\documentclass[11pt]{article}

\usepackage[a4paper, top=2.5cm, bottom=2.5cm, left=3cm, right=3cm]{geometry}
\usepackage{xcolor}
\usepackage{amsmath}
\usepackage{amssymb}
\usepackage{amsfonts}
\usepackage{graphicx}
\usepackage{subcaption}
\usepackage{float}
\usepackage{soul}
\usepackage[unicode]{hyperref}
\usepackage{listings}
\usepackage{booktabs}
\usepackage[utf8]{inputenc}
\usepackage{cite}
\usepackage[T1]{fontenc}
\usepackage{lmodern}

\usepackage[affil-it]{authblk}

\definecolor{codegray}{rgb}{0.95,0.95,0.95}
\newcommand{\ytilde}{\tilde{y}}
\newcommand{\Adec}{A_{\rm dec}}
\newcommand{\Bgrow}{B_{\rm grow}}
\newcommand{\PINN}{\textsc{pinn}}
\newcommand{\R}{\mathbb{R}}
\newcommand{\atilde}{\tilde a}
\newcommand{\btilde}{\tilde b}
\newcommand{\ctilde}{\tilde c}
\newcommand{\Nex}{N_{\varepsilon}}
\newcommand{\lint}{\ell_{\rm int}}

\newcommand{\aeps}{a_{\varepsilon}}
\newcommand{\beps}{b_{\varepsilon}}
\newcommand{\ceps}{c_{\varepsilon}}

\long\def\chg#1{\textcolor{red}{#1}}            
\definecolor{chgblue}{rgb}{0.0,0.25,0.85}
\long\def\chgB#1{\textcolor{chgblue}{#1}}       
\definecolor{todoorange}{rgb}{0.85,0.45,0.0}
\long\def\todo#1{\textcolor{todoorange}{[\textsf{TO DO:} #1]}}

\definecolor{chgpurple}{rgb}{0.55,0.10,0.60}

\usepackage[normalem]{ulem}                     

\definecolor{mariateal}{rgb}{0.0,0.50,0.50}
\newcommand{\maria}[1]{\textcolor{mariateal}{#1}}

\title{Black Hole Love Numbers with\\
       Physics-Informed Neural Networks}
\author[1,2]{Mar\'ia J.\ Rodr\'iguez%
  \thanks{\texttt{majo.rodriguez.b@gmail.com}}}
\author[1]{Luis Fernando Temoche%
  \thanks{\texttt{l.f.temoche@usu.edu}}}
\author[3]{Pavlos Protopapas%
  \thanks{\texttt{pavlos@seas.harvard.edu}}}

\affil[1]{Department of Physics, Utah State University,
          4415 Old Main Hill Road, UT 84322, USA}
\affil[2]{Instituto de F\'isica Te\'orica UAM/CSIC,
          Universidad Aut\'onoma de Madrid,
          13-15 Calle Nicol\'as Cabrera, 28049 Madrid, Spain}
\affil[3]{John A.\ Paulson School of Engineering and Applied Sciences, Harvard University, Cambridge, MA 02138, USA}
\date{}
\begin{document}

\maketitle

\begin{abstract}
We develop and validate a physics-informed neural network (PINN) framework for computing the tidal Love numbers of black holes directly from their perturbation equations. We consider static perturbations of the Kerr black hole, whose exact hypergeometric solutions provide a controlled benchmark for both the numerical solver and the extraction of the physical response. The framework combines exact enforcement of the horizon boundary conditions with an \(H^2\)-Sobolev loss that incorporates derivatives of the differential-equation residual. The Love number is then obtained by extracting the asymptotic coefficients of the learned solution, without using the analytic connection formula. Across the tested scalar, vector, and tensor perturbations, the method recovers the finite dissipative response with relative errors below \(10^{-5}\). Controlled comparisons show that the higher-derivative loss terms improve solution accuracy by orders of magnitude, while independent tests of the extraction procedure distinguish solver errors from errors in recovering the response coefficients. The conservative response, which vanishes analytically, is constrained by the numerical error floor. These results demonstrate that PINNs can accurately compute black-hole tidal response and establish a validated framework for future extensions to frequency-dependent and non-separable perturbation problems.
\end{abstract}

\newpage

\tableofcontents

\section{Introduction}
Physical observables are often determined indirectly from solutions of
differential equations: the equation determines a function, and the quantity
of interest is read off from its boundary or asymptotic
behaviour. Examples include response coefficients, scattering amplitudes,
eigenvalues, and quasinormal frequencies. The accuracy of the observable
therefore depends not only on the accuracy of the numerical solution, but
also on the procedure used to extract it.

The extraction step becomes important when the observable is small or vanishes
exactly. Such a quantity is typically a subdominant
coefficient in an asymptotic expansion, or the result of a cancellation between much larger
contributions. Errors in the extracted quantity can then greatly exceed
errors in the underlying solution.

Tidal Love numbers characterise
the linear response of a compact object to an external gravitational field
and are relevant to gravitational-wave observations and to tests of
strong-field gravity. For black holes the Love number is a complex
coefficient: its real, conservative part measures the deformability of the horizon under
an external perturbation, and its imaginary, dissipative part the rate of energy dissipation.
Both matter observationally in the inspiral of a binary, when each object is tidally deformed
by its companion. For four-dimensional black holes in
general relativity the conservative Love numbers vanish identically in the
static limit \cite{Kol:2011vg, Gurlebeck:2016rii, Hui:2020xxx, Riva:2023rcm,
Hadad:2024lsf}, while the response of a spinning black hole also contains a
finite dissipative part \cite{LeTiec:2020bos, Chia:2020yla,
Charalambous:2021mea, Perry:2023wmm}. The vanishing of the conservative part is explained
by a set of hidden conformal symmetries \cite{Hui:2021vcv, Charalambous:2021kcz,
Charalambous:2022rre, Hui:2022vbh, BenAchour:2022uqo, Katagiri:2022vyz,
Kehagias:2022ndy, Guevara:2023wlr, Cvetic:2024dvn, Gray:2024qys}. Both a vanishing and a finite response
coefficient are therefore available within the same perturbation problem, which makes Love
numbers a clean test of this effect.

In higher dimensions the pattern is richer and the static tidal Love
numbers do not universally vanish \cite{Charalambous:2023jgq,
Rodriguez:2023xjd, Charalambous:2024tdj, Glazer:2024eyi,
Charalambous:2025ekl, Cvetic:2026wht}. For a
comprehensive review of tidal deformations of compact objects see
\cite{Chakraborty:2026qru, Rodriguez:2026iot}.

The calculation of tidal Love numbers has received considerable attention in
recent years. Approaches include direct numerical methods, effective field
theory \cite{Ivanov:2022hlo, Ivanov:2022qqt, Saketh:2023bul, Barura:2024uog, Glazer:2024eyi, 
Caron-Huot:2025tlq, Combaluzier--Szteinsznaider:2025eoc, Barbosa:2025uau,  Solon:2026ubm, Barbosa:2026qcv, Elkhidir:2026lgp},
master-equation analyses via perturbative schemes such as the MST formalism
\cite{Perry:2024vwz} or connection-formula implementations
\cite{Rodriguez:2025ala}, and dualities such as the AGT correspondence
\cite{Bonelli:2021uvf, Solodukhin:2026vtz}.
{These methods are accurate within their perturbative domain of
validity.}
The traditional route to the computation of the Love number involves the
linear transformation formula relating the horizon and asymptotic boundary
expansions, expressed in closed form via products of Gamma functions,
\begin{equation} \label{eq:Love}
k_\ell = \frac{\Gamma(1+\ell-s)\,\Gamma(1+\ell + 2 i m \gamma)\,
               \Gamma(-1-2\ell)}
              {{\Gamma(2 i m \gamma-\ell)}\,{\Gamma(-\ell-s)}\,
               \Gamma(1 + 2\ell)}\,.
\end{equation}
The coefficients appearing in the corresponding expression are defined and
discussed in more detail later in the text.
{This approach is analytic, but at the integer angular numbers that arise
in physical applications the two Gamma functions $\Gamma(-1-2\ell)$ in the numerator and
$\Gamma(-\ell-s)$ in the denominator both sit at poles, while every other factor stays finite
provided $m\gamma\neq0$ (for $m\gamma=0$ the denominator factor $\Gamma(-\ell)$ is singular
as well, and $k_\ell$ vanishes).
The Love number is therefore defined only through the cancellation of these
two divergences, and the limit must be approached rather than evaluated
directly.}

In this work we compute the Love numbers without the connection formula, with a
fully numerical approach based on Physics-Informed Neural Networks (\PINN s)
\cite{raissi2019physics}. The \PINN\ solves the perturbation equation, and the Love number
is then extracted from the asymptotic coefficients of the learned solution. The calculation
therefore has exactly the two steps described above, and because the exact solution is known,
the error of each step can be measured separately. \PINN s have previously been applied to quasinormal
modes \cite{Luna:2022rql, Cornell:2022enn, Patel:2024iij} for the
Schwarzschild and Kerr cases, and generalized solvers for the Regge--Wheeler
and Teukolsky equations have been developed \cite{Cornell:2024azz}.
Extensions cover asymptotically dS and AdS solutions \cite{Ovgun:2019yor,
Ncube:2021jfu}, higher dimensions \cite{Lobos:2024fzj} and modified gravity
\cite{Haddou:2023eqp, Luna:2024spo}.

Neural-network solvers are also relevant in a setting where the standard
methods do not apply at all. \PINN s do not require separability.
The Teukolsky construction,
and with it the expansions and connection formulae, rests on
the separation of the perturbation equation into radial and angular parts, a
structure special to Kerr and to a handful of related backgrounds. For
modified-gravity black holes, for numerically constructed stationary
solutions, and for rotating solutions in higher dimensions, no exact separation is known in
general, in particular for gravitational perturbations at finite spin, and the tidal response
is known mostly in the non-rotating or slowly rotating
limits\maria{~\cite{Cardoso:2017cfl,DeLuca:2022tkm,Barura:2024uog}}.
A residual-minimising solver is not tied to separability: the residual loss carries over to a
partial differential equation in several variables, although the boundary treatment and the
extraction of the response would have to be adapted.
This has recently been demonstrated for the quasinormal-mode problem, where
the two-dimensional hyperboloidal Teukolsky equation was solved directly,
without separation, as a prototype for beyond-Kerr backgrounds
\cite{FerrerSanchez:2026qnm}. Spectral methods that avoid separation also
exist and are substantially more accurate \cite{Chung:2023wkd,
BlazquezSalcedo:2023hwg}; the case for a \PINN\ rests on flexibility and on
requiring little analytical pre-processing, not on precision.

The \PINN\ applications cited above are mostly quasinormal-mode calculations, which are
eigenvalue problems. The
boundary conditions select a discrete frequency, and the accuracy of that
frequency is the accuracy of the result. A tidal response coefficient is a
connection problem of a different character. The frequency is fixed (zero in the static case
studied here), a
solution exists for every parameter value, and the observable is a ratio of
two coefficients in the expansion of that solution about the boundary, one of
which multiplies a strongly suppressed local behaviour. The accuracy of the
solution and the accuracy of the observable are then distinct questions, and
the second does not follow from the first. That gap is the subject of this
paper.

The Kerr black hole tidal Love number problem is well suited to a controlled
numerical study. The full frequency-dependent perturbation equation is of
confluent-Heun type \cite{ronveaux1995heun}, while in the static limit
$\omega=0$ it reduces to a hypergeometric equation. In this paper, we use the
static equation as a controlled test case, leaving the non-static case for a
follow-up study. Because the static equation admits an exact hypergeometric
solution, errors in solving the differential equation can be separated from
errors introduced by extracting the physical response. Instead of evaluating
the analytic connection formula, we train a neural network to represent the
solution regular at the horizon directly and extract the asymptotic
coefficients $\Adec$ and $\Bgrow$ by fitting the solution to the two independent solutions
defined at the asymptotic boundary.

Our primary goal is to develop a highly accurate solver for the differential
equation governing black-hole perturbations and, at the same time, a reliable
procedure for extracting the associated physical observable, the Love number.
The static problem provides an ideal controlled setting in which to develop
and test both components: the exact solution is known, the relevant
asymptotic data can be extracted explicitly, and the accuracy of the
resulting Love number can be assessed independently.
The object of study is accordingly the extraction step as much as the
solver. The static Love number is used because its exact value is known,
which is what makes it possible to separate solver error from extraction
error. The conclusions drawn about that separation apply to any solver whose
output is sampled and projected onto an asymptotic basis. For the solver, we use a weighted
\(H^2\)-Sobolev norm for
the training loss. The loss penalizes the differential-equation residual and
its first two derivatives, rather than the residual alone, providing greater
control over the local behavior of the solution and its derivatives.
The weight $w=1/[z(1-z)]$ diverges at the two singular endpoints and so
penalises the residual most strongly where the Frobenius structure of the
solution is delicate.

A small differential-equation residual alone, however, does not guarantee an
accurate physical observable. We therefore implement and analyze different
methods for reliably extracting physical information encoded in the
asymptotic behavior of the solution. This distinction is important here
because the Love number is obtained from a delicate combination of terms in
the near-boundary expansion, with \(\Adec\) multiplying a high-order
asymptotic contribution.

Since the analytical static solution is known, we can disentangle errors in the
neural-network solver from those introduced by the extraction procedure. We
apply the same extraction method to both the Frobenius solution and the
trained network. Errors common to the two calculations arise from the
extraction procedure, whereas additional errors in the network result
originate from the numerical solver.

The static problem is also the natural first step toward the more general,
frequency-dependent problem. Our aim in future work is to retain the solver
and observable-extraction framework developed here and subsequently extend it
to nonzero \(\omega\), where the analytic structure is substantially more
involved. Beyond the black-hole application, the same framework provides a
general approach to second-order ODEs with singular points. In particular,
from the perspective of singular-point classification, the present equation
is the hypergeometric member of the broader class containing the general Heun
equation and its confluent limits. Related approaches have recently been
explored \cite{Cayuso:2024jau, bonidifferential}, but, to the best of
our knowledge, have been restricted to real parameters and real-valued
solutions. Our solver instead treats the hypergeometric equation with
complex parameters and complex-valued solutions.

The paper is organized as follows. Section~\ref{sec:hyper} reviews the hypergeometric equation, introduces a regulator that lifts the degeneracy of the connection problem, and presents a field redefinition that regularizes the behavior at \(z=1\). It also establishes that the static Love number is identical for spin weights \(s\) and \(-s\). Section~\ref{sec:realsystem} recasts the resulting equation as a coupled system of two real ODEs. Section~\ref{sec:pinn} introduces the \PINN\ architecture, the weighted \(H^2\) loss, and the training procedure, explaining how the horizon boundary condition is imposed and motivating our choice to solve the equation at spin weight \(-s\). Section~\ref{sec:amplification} analyzes how errors propagate from the loss function to the solution and the extracted coefficients, and presents the trained solution for our flagship configuration. Section~\ref{sec:extraction} introduces eight methods for extracting \(\Adec\) and \(\Bgrow\) and presents the corresponding numerical results. Section~\ref{sec:localfail} provides a closed-form explanation of why one family of extraction methods necessarily fails, while Section~\ref{sec:convention} examines the exponent convention that governs the accuracy of the entire computational pipeline. The subsequent subsections investigate the role of the Sobolev loss, the dependence of coefficient errors on the regulator \(\varepsilon\), the resulting physical Love number, and the behavior across angular modes \((s,\lint,m)\). Finally, Section~\ref{sec:discussion} summarizes our conclusions and outlines directions for future work, while Appendix~\ref{app:teukolsky} derives the hypergeometric form of the static Teukolsky equation.

\section{Set-up}
\label{sec:hyper}
Our starting point is the radial Teukolsky equation \cite{Teukolsky:1972my,
Teukolsky:1973ha, Press:1973zz, Teukolsky:1974yv} governing perturbations of
the Kerr background, evaluated in the static limit $\omega=0$. Throughout we
work in units $M=1/2$, with $M$ the black-hole mass, so that the horizon
radii are
\begin{equation}\label{eq:horizons}
  r_\pm=\tfrac{1}{2}\bigl(1\pm\sqrt{1-4a_{\rm Kerr}^{2}}\bigr),
\end{equation}
and spin parameter $a_{\rm Kerr}\le 1/2$, with extremality occurring at
$a_{\rm Kerr}=M=1/2$. The details of the reduction are given in
Appendix~\ref{app:teukolsky}; here we record the result. With the
compactified radial variable
\begin{equation}\label{eq:zvar}
  z=\frac{r-r_+}{r-r_-},
\end{equation}
which maps the black hole horizon $r=r_+\mapsto z=0$, the asymptotic boundary
$r=\infty\mapsto z=1$ and Cauchy horizon $r=r_-\mapsto z=\infty$, and with
the substitution
\begin{equation}\label{eq:Rhat_to_y_main}
  \hat R_s(r) = z^{\,-s/2+im\gamma}\,(1-z)^{\,\ell+1}\,y(z),
\end{equation}
the static radial equation becomes the hypergeometric equation
\cite{Perry:2023wmm}
\begin{equation}\label{eq:hyper}
  z(1-z)\,y'' + \bigl[c-(a+b+1)z\bigr]\,y' - ab\,y \;=\; 0,
\end{equation}
with parameters
\begin{equation}\label{eq:abc}
  a = 1+\ell-s, \qquad
  b = 1+\ell + 2 i m \gamma, \qquad
  c = 1-s + 2 i m \gamma .
\end{equation}
Here $\ell$ and $m$ are the angular and azimuthal numbers and $s$ is the spin
weight of the perturbing field, constrained by $|s|\le \ell$ and
$|m|\le \ell$. The black-hole spin enters only through the single combination
\begin{equation}\label{eq:gamma}
  \gamma \;=\; \frac{a_{\rm Kerr}}{r_+-r_-} .
\end{equation}

Two features of \eqref{eq:abc} will be important in what follows. First, $b$
and $c$ are complex whenever $m\gamma\neq0$, while $b-c=\ell+s$ is an
integer. The complex parameters make $y$ complex-valued, which is why the
\PINN\ of Section~\ref{sec:realsystem} is trained on a coupled system for the
real and imaginary parts. They also give rise to the dissipative (imaginary)
part of the Love number: when $m\gamma=0$, all parameters are real and
$k_\ell$ is real. The integer $b-c$ makes the Gamma-ratio
formula~\eqref{eq:AB_gamma} indeterminate at physical parameter values, since
$1/\Gamma(c-b)$ vanishes exactly where $\Gamma(c-a-b)$ has a pole.
It has a second consequence. With
$c-b$ a non-positive integer the hypergeometric series~\eqref{eq:euler} for
$\ytilde$ terminates, so at integer angular number $\ytilde$ is a polynomial
of degree $\ell+s$. Since $|s|\le\ell$ implies $\ell+s<2\ell+1$, both local
solutions at $z=1$ are then analytic and no logarithm appears. This is the
degenerate case \emph{without} a logarithmic branch, and it is the same fact
that makes the conservative response vanish.

Second, the combination
\begin{align}\label{eq:Ndef}
  a+b-c \;=\; 2\ell+1 \;\equiv\; N
\end{align}
is the difference of the Frobenius exponents at $z=1$. Recall that the two
Frobenius behaviors there are $(1-z)^0$ and $(1-z)^{c-a-b}$, corresponding in
the $r$-coordinates to the $r^{-(\ell+1)}$ response and the $r^\ell$ tidal
field, respectively (Appendix~\ref{app:teukolsky}). For integer $\ell$, this
difference is the positive integer $2\ell+1$, placing the equation in the
\emph{degenerate} (resonant) case of the hypergeometric family.

Rather than working directly with the degenerate form of the connection
formula, we regulate the exponent difference by analytically continuing the
angular index,
\begin{equation}\label{eq:regulator}
\ell \;\longrightarrow\; \lint+\varepsilon,
\qquad
\varepsilon\neq0,
\end{equation}
where $\lint\in\mathbb{Z}$ denotes the physical angular number.
The corresponding hypergeometric parameters become
\begin{equation}\label{eq:abceps}
\aeps=1+\lint+\varepsilon-s,
\qquad
\beps=1+\lint+\varepsilon+2im\gamma,
\qquad
\ceps = c =1-s+2im\gamma,
\end{equation}
the last because $c$ does not depend on $\ell$. The exponent difference
at $z=1$ becomes
\begin{equation}\label{eq:Nex}
\aeps+\beps-\ceps \;=\; 2(\lint+\varepsilon)+1
  \;\equiv\; \Nex \;=\; N+2\varepsilon .
\end{equation}
We reserve $\Nex$ for this real valued, non-integer exponent difference and
$N=2\lint+1$ for its integer part. The two coincide as $\varepsilon\to0$, but
they differ at $\mathcal{O}(\varepsilon)$, which is the same order as the
quantity ultimately being measured.

For nonzero $\varepsilon$, the exponent difference is noninteger and the
standard Kummer connection formula may be applied without encountering the
poles associated with the resonant case. We therefore first perform the
analytic continuation and the expansion about $z=1$ at finite $\varepsilon$,
and only subsequently take the limit $\varepsilon\to0$. The tidal 
response coefficients can therefore be extracted before taking the resonant
limit, after which the physical static Love number is obtained from the
resulting finite ratio.
The continuation is a formal device and carries no physical content. For non-integers the angular equation no longer admits solutions regular on the
full domain, and we use \eqref{eq:regulator} solely to define the
$\varepsilon\to0$ limit of the radial connection problem, which is where the
physical statement resides.

Of the two solutions to the second order differential equation in
Eq.~\eqref{eq:hyper}, the regularity condition at the horizon ($z=0$) selects
the regular branch ${}_2F_1(\aeps,\beps;\ceps;z)$. Expanding this
horizon-regular solution about $z=1$ using the standard linear transformation
formula\footnote{Known in the mathematics literature as the Kummer connection
formula \cite{abramowitz1948handbook, bateman1953higher}} gives
\begin{align}\label{eq:connection}
  y={}&
  \Adec\, {}_2F_1(\aeps,\beps;\aeps+\beps-\ceps+1;1-z)
  \nonumber\\
  &+\Bgrow\, (1-z)^{\ceps-\aeps-\beps}\,
  {}_2F_1(\ceps-\aeps,\ceps-\beps;\ceps-\aeps-\beps+1;1-z),
\end{align}
where the connection coefficients $\Adec$ and $\Bgrow$ are determined by the
hypergeometric transformation and are given by
\begin{equation}\label{eq:AB_gamma}
  \Adec = \frac{\Gamma(\ceps)\,\Gamma(\ceps-\aeps-\beps)}
               {\Gamma(\ceps-\aeps)\,\Gamma(\ceps-\beps)},
  \qquad
  \Bgrow = \frac{\Gamma(\ceps)\,\Gamma(\aeps+\beps-\ceps)}
               {\Gamma(\aeps)\,\Gamma(\beps)} .
\end{equation}
Note that $\Bgrow$ remains finite and non-zero as $\varepsilon\to0$: its
argument $\aeps+\beps-\ceps=\Nex$ is positive, and neither $\Gamma(\aeps)$
nor $\Gamma(\beps)$ has a pole there. The entire singular structure resides
in $\Adec$, through the ratio
$\Gamma(\ceps-\aeps-\beps)/\Gamma(\ceps-\beps)$ of two divergent factors.
The static Love number is then \cite{Perry:2023wmm}
\begin{equation}\label{eq:kl}
  k_\ell \;=\; \\lim_{\varepsilon \rightarrow 0}\frac{\Adec}{\Bgrow},
\end{equation}
which agrees with Eq.~\eqref{eq:Love}. This analytic procedure motivates a
complementary numerical approach with \PINN s. Before Eq.~\eqref{eq:hyper}
can be handed to a numerical solver, however, several considerations must be
addressed, since the aim is a solver accurate enough to resolve the
subleading coefficient $\Adec$ and not merely the solution. First, the
boundary conditions must be imposed on a function that stays bounded at both
singular points, whereas $y$ itself grows as $(1-z)^{-\Nex}$ at $z=1$.
Second, a noise-free reference solution is needed against which the solver
and the extraction of $\Adec$ and $\Bgrow$ can be tested independently.
Third, the parameters $\beps$ and $\ceps$ are complex, so the equation is
complex-valued and the solver must handle a coupled system for its real and
imaginary parts. We take these in turn.

The spin weight enters Eq.~\eqref{eq:Love} only through the
factor $\Gamma(1+\ell-s)/\Gamma(-\ell-s)$. Applying
$\Gamma(x)\Gamma(1-x)=\pi/\sin\pi x$ to both Gamma functions, the ratio of
this factor at $-s$ to its value at $s$ is
$\sin\pi(1+\ell-s)/\sin\pi(1+\ell+s)$, which equals one for integer $s$ and
any $\ell$, including $\ell=\lint+\varepsilon$. The static Love number is
therefore the same whether it is computed from the equation with spin weight
$s$ or $-s$, in line with the Teukolsky--Starobinsky identities relating the
two \cite{Teukolsky:1974yv}. The boundary conditions are also unchanged:
with the prefactor of Eq.~\eqref{eq:Rhat_to_y_main}, the ingoing solution
$R_s\sim\Delta^{-s}e^{-ikr_*}$ corresponds to the exponent $0$ of $y$ at
$z=0$ for every $s$, and the exponents at $z=1$ do not depend on $s$
(Appendix~\ref{app:teukolsky}). We make use of this in later sections.

\subsection{Boundary conditions}

To impose boundary conditions appropriate for tidal deformations and to
extract Love numbers from a smooth function, we adopt the field redefinition
\begin{align}
  \ytilde(z) \;=\; (1-z)^{\Nex}\, y(z) ,
  \label{eq:difftilde}
\end{align}
with $\Nex=2(\lint+\varepsilon)+1$ the exponent difference at $z=1$, Eq.~\eqref{eq:Nex}, after which $\ytilde$ is regular at both singular points. The exponent
here is not a matter of choice. Under $y\mapsto(1-z)^{\lambda}y$ the
exponents at $z=1$ shift from $\{0,\,\ceps-\aeps-\beps\}$ to
$\{\lambda,\,\ceps-\aeps-\beps+\lambda\}$, while those at $z=0$ are
untouched. For the transformed equation to remain hypergeometric one exponent
at $z=1$ must vanish, which forces $\lambda=0$ or
$\lambda=\aeps+\beps-\ceps=\Nex$.
Equation~\eqref{eq:difftilde} is thus the unique non-trivial rescaling within
this family that preserves the hypergeometric class; any other exponent, the
integer $N$ included.
Under \eqref{eq:difftilde} Eq.~\eqref{eq:hyper} retains hypergeometric form,
\begin{align}\label{eq:hyper_tilde}
  z(1-z)\,\ytilde'' + \bigl[\ctilde - (\atilde + \btilde + 1) z\bigr]
  \ytilde' - \atilde \btilde\, \ytilde \;=\; 0,
\end{align}
with shifted parameters
\begin{align}\label{eq:tildeparams}
  \atilde = \aeps - \Nex = \ceps-\beps, \qquad
  \btilde = \beps - \Nex = \ceps-\aeps, \qquad
  \ctilde = \ceps ,
\end{align}
so that $\atilde+\btilde-\ctilde = \ceps-\aeps-\beps = -\Nex$. Equivalently, by Euler's transformation,
\begin{equation}\label{eq:euler}
\ytilde(z) \;=\; {}_2F_1(\ceps-\aeps,\,\ceps-\beps;\,\ceps;\,z),
\end{equation}
which is the closed form against which every numerical statement below
can be verified.
The tilded solution $\ytilde$ is regular at both singular points, making it a
convenient variable for the numerical treatment. The horizon boundary
condition is simply
\begin{align}\label{eq:bc_horizon}
\boxed{\ytilde(0)=1}\,.
\end{align}
Near spatial infinity, $z=1$, we expand $\ytilde$ in the two local solutions
of the tilded hypergeometric equation,
\begin{align}\label{eq:expansion_at_1}
\ytilde(z)
=\Adec\,f_1(z)+\Bgrow\,f_2(z),
\qquad
f_1(z)\sim(1-z)^{\Nex},
\quad
f_2(1)=1,
\end{align}
where
\begin{align}
f_1(z)
&=(1-z)^{\Nex}\,
{}_2F_1\!\left(\ctilde-\atilde,\;\ctilde-\btilde;\;\Nex+1;\;1-z\right),\label{eq:basis_def}\\
f_2(z)
&={}_2F_1\!\left(\atilde,\;\btilde;\;1-\Nex;\;1-z\right). \label{eq:basis_def2}
\end{align}
The coefficients $\Adec$ and $\Bgrow$ therefore characterize the two
asymptotic sectors and determine the corresponding response-to-tidal ratio.
We note for later use that $f_1$ and $f_2$ need not be evaluated as
hypergeometric functions at all. Substituting $w=1-z$ in
Eq.~\eqref{eq:hyper_tilde} returns a hypergeometric equation in $w$ with
parameters $(\atilde,\btilde,\atilde+\btilde+1-\ctilde)$, so the exponents
$\{0,\Nex\}$ and the two series may be generated directly from the
coefficients of the differential equation by Frobenius recursion,
\begin{align}
f_1&=w^{\Nex}\!\sum_{k\ge0} h_k w^k,\quad
\frac{h_{k+1}}{h_k}=\frac{(k+\atilde+\Nex)(k+\btilde+\Nex)}
                          {(k+1)(k+1+\Nex)} \label{eq:basis_frobenius}\\
                          f_2&=\sum_{k\ge0} g_k w^k,\quad \label{eq:basis_frobenius2}
\frac{g_{k+1}}{g_k}=\frac{(k+\atilde)(k+\btilde)}{(k+1)(k+1-\Nex)}\,.
\end{align}
Note that the Frobenius recursion only fixes the coefficient ratios so the leading coefficients $h_0$ and $g_0$ are free. We adopt the standard choice $h_0 = g_0 = 1$, 
The Love number identification  therefore requires only
the \emph{local} structure at $z=1$, and not the global connection formula.

Condition~\eqref{eq:bc_horizon} alone does not fix the solution, because the
horizon is a regular singular point. In Eq.~\eqref{eq:hyper_tilde} the
coefficient $z(1-z)$ of the highest derivative vanishes at $z=0$, so the
equation degenerates into an \emph{algebraic} relation between $\ytilde(0)$
and $\ytilde'(0)$:
\begin{equation}\label{eq:indicial}
  \ctilde\,\ytilde'(0) \;-\; \atilde\,\btilde\,\ytilde(0) \;=\; 0 .
\end{equation}
Combining Eq.~\eqref{eq:bc_horizon} with Eq.~\eqref{eq:indicial} fixes the
derivative,
\begin{equation}\label{eq:yp0}
  \boxed{\;\ytilde'(0) \;=\; \frac{\atilde\,\btilde}{\ctilde}\;}
\end{equation}
We emphasize that Eq.~\eqref{eq:yp0} is \emph{not} an additional physical
input: it is forced by the ODE once one imposes Eq.~\eqref{eq:bc_horizon} and
demands that $\ytilde$ be analytic at the horizon.


\subsection{Decomposition into real and imaginary parts}
\label{sec:realsystem}
Because
some of the coefficients carry the imaginary part $2im\gamma$, the tilded
equation~\eqref{eq:hyper_tilde} has complex coefficients and $\ytilde$ is a
complex-valued function of a real variable. A neural network with real
weights and real activations cannot represent this directly, and automatic
differentiation frameworks operate natively on real tensors. We therefore
split the equation into its real and imaginary parts and solve the resulting
coupled system for two real functions on $[0,1]$. The two components are coupled
precisely through the imaginary parts of the parameters, which is what
carries the dissipative response.

We write
\begin{align}
  \ytilde(z) = u(z) + i\, v(z),
  \qquad \atilde = a_r + i a_i, \quad
        \btilde = b_r + i b_i, \quad
        \ctilde = c_r + i c_i,
\end{align}
and define
\begin{align}
  \alpha &= a_r b_r - a_i b_i, \qquad
  \beta  = a_r b_i + a_i b_r, \\
  P(z)   &= c_r - (a_r + b_r + 1) z, \qquad
  Q(z)   = c_i - (a_i + b_i) z.
\end{align}
Here $\alpha$ and $\beta$ are the real and imaginary parts of the
product $\atilde\btilde$, while $P$ and $Q$ are those of the linear
coefficient $\ctilde-(\atilde+\btilde+1)z$.
Substituting into Eq.~\eqref{eq:hyper_tilde} and equating real and imaginary
parts gives the coupled system
\begin{equation}\label{eq:system}
  \boxed{
  \begin{aligned}
    z(1-z)\,u'' + P(z)\,u' - Q(z)\,v' - \alpha\, u + \beta\, v &= 0, \\[4pt]
    z(1-z)\,v'' + P(z)\,v' + Q(z)\,u' - \alpha\, v - \beta\, u &= 0,
  \end{aligned}}
\end{equation}
with boundary conditions $u(0) = 1$, $v(0) = 0$.
The two equations share the same second-order operator and differ only
in the sign with which $Q$ and $\beta$ couple them; setting $m\gamma=0$ makes
$Q=\beta=0$, the system decouples, and $v\equiv0$ solves the imaginary part
identically, recovering the real problem noted below Eq.~\eqref{eq:gamma}.
This is the system the \PINN\ is trained to solve.

\section{Physics-Informed Neural Network architecture}
\label{sec:pinn}
Having reduced the problem to the real system~\eqref{eq:system} with the
boundary data~\eqref{eq:bc_horizon} and \eqref{eq:yp0}, we now describe the
neural-network approach to solving it. A \PINN\ represents the unknown
solution by a feed-forward network and trains it by minimising the residual
of the differential equation itself, evaluated at collocation points drawn
from the domain; no solution data are required. Two ingredients are specific
to the present problem: how the boundary
conditions at the regular singular point $z=0$ are imposed, and what norm of
the residual is minimised.

A unique solution of a differential equation is singled out by its boundary
conditions, and for \PINN s these may be imposed in two ways
\cite{alkhadhr2023wave, sheikholeslami2025physics}. Under \emph{soft}
enforcement the boundary conditions enter as additional terms in the loss competing with the bulk residual during
optimisation. Under \emph{hard} enforcement the conditions are built into the
architecture, so they hold identically for every value of the network
parameters, either through a trial function that satisfies them by construction \cite{Lagaris:1997at, chen2020neurodiffeq,liu2025recent} or by constraining the network itself \cite{mattheakis2019physical}. We use hard enforcement throughout, for reasons that are specific
to this problem and are set out below.

\subsection{Weighted \texorpdfstring{$H^{2}$}{H2}-Sobolev loss}

The standard \PINN\ loss penalises the differential-equation residual, but does not explicitly penalise its derivatives.
Define the real and imaginary parts of the ODE residual from \eqref{eq:system} as
\begin{align}
  R_u(z) &= z(1-z)\,u'' + P u' - Q v' - \alpha u + \beta v, \label{eqn:ODEred1}\\
  R_v(z) &= z(1-z)\,v'' + P v' + Q u' - \alpha v - \beta u. \label{eqn:ODEred2}
\end{align}
Standard \PINN\ formulations minimize $\int (R_u^2 + R_v^2)\,dz$. 
Since we are solving a second-order equation and the Love number $k_\ell$ is obtained from the expansion coefficients, accurate recovery of the solution and its derivatives is important. We therefore also penalise the first and second derivatives of the residual. This gives an $H^2$-Sobolev loss: a loss that combines the squared residual with its squared derivatives up to second order \cite{czarnecki2017sobolev, son2021sobolev}. For background on Sobolev spaces and weak derivatives, see \cite{kreyszig1991introductory, adams2003sobolev, brezis2011functional, evans2022partial}. Independently of these derivative penalties, we introduce an endpoint weight $w(z)$ to give greater emphasis to the regions near the two singular boundaries. The resulting weighted loss is
\begin{equation}\label{eq:loss}
  \mathcal{L}(\theta) \;=\;
  \int_{z_{\min}}^{z_{\max}} w(z)\Bigl[
      \bigl(R_u^2 + R_v^2\bigr)
      + \lambda_1\bigl((R_u')^2 + (R_v')^2\bigr)
      + \lambda_2\bigl((R_u'')^2 + (R_v'')^2\bigr)
  \Bigr]\,dz ,
\end{equation}
with weight
\begin{equation}\label{eq:weight}
  w(z) \;=\; \frac{1}{z(1-z)} ,
\end{equation}
and Sobolev coefficients $\lambda_1 = 10^{-1}$, $\lambda_2 = 10^{-2}$. {Here primes denote derivatives of the residuals with respect to $z$. Their explicit expressions are given in Appendix~\ref{app:residual_derivatives}.}

\definecolor{fillgreen}{rgb}{0.0,0.50,0.20}
\long\def\runval#1{{\color{fillgreen}#1}}
\long\def\chg#1{{\color{red}#1}}
\long\def\chgB#1{{\color{chgblue}#1}}
\long\def\todo#1{{\color{todoorange}[\textsf{TO DO:} #1]}}

We comment below on several choices made in the construction of the loss.

{In practice the integral is replaced by the empirical mean of the weighted integrand over $400$ collocation points, drawn uniformly on $[z_{\min},z_{\max}]=[0.01,0.99]$ and redrawn at each optimisation step.}

{We choose $w(z)=1/[z(1-z)]$, the reciprocal of the coefficient of $\ytilde''$ in Eq.~\eqref{eq:hyper_tilde}, to increase the penalty near the horizon and the asymptotic boundary. These are the regions where the boundary conditions and expansion coefficients are determined. This is a numerical weighting choice, rather than a requirement of a Sobolev loss. Its effect is tested separately from that of the derivative penalties in Table~\ref{tab:crossed}. Because the weight diverges at both endpoints, we evaluate the loss on the truncated interval $[0.01,0.99]$, where it remains finite.}

{We first derive the first and second derivatives of the residuals analytically by applying the product rule to Eqs.~\eqref{eqn:ODEred1}--\eqref{eqn:ODEred2}, using the explicit linear forms of $P$ and $Q$. The resulting expressions are given in Appendix~\ref{app:residual_derivatives}. We evaluate the derivatives of the network outputs $u$ and $v$ appearing in these expressions using automatic differentiation. The standard residual requires network derivatives up to second order; the first- and second-derivative Sobolev terms require network derivatives up to third and fourth order, respectively.}

The first term alone in \eqref{eq:loss} without the Sobolev terms
admits residuals that are small in mean but
oscillatory. Including $(R')^2$ suppresses such oscillations, and $(R'')^2$
further enforces smoothness of the residual itself. We use $\lambda_1=0.1$
and $\lambda_2=0.01$ as a conservative hierarchy in which each
higher-derivative penalty is subdominant to the previous.
These values were fixed before the sensitivity scan, by a short trial on the
$s=\lint=m=2$ benchmark at $\varepsilon=10^{-2}$ as an order-of-magnitude
choice making each penalty roughly ten times smaller than the previous one.
The results depend only weakly on this choice (Section~\ref{sec:sobolev}, Table~\ref{tab:lambda_sens}).

Our training procedure uses Adam followed by an L-BFGS refinement stage. For the initial benchmark, Adam alone achieves the reported accuracy, and the L-BFGS stage is omitted (Section~\ref{sec:training}). Both stages are used for the angular-mode scan (Section~\ref{sec:lmscan}). The Adam stage uses the optimizer {Adam \cite{kingma2015adam}} at base learning rate $10^{-3}$, using $400$ collocation
points drawn uniformly at each epoch.
During training we keep the network with the lowest \emph{validation
loss}, the same loss evaluated on a fixed grid of $4000$ equally spaced points
in $z\in[0.01,0.99]$ that are not used for training, rather than the loss on the randomly drawn training points, which varies from epoch to epoch.
{For the angular-mode scan,} the Adam stage ($750$ epochs) is followed by L-BFGS. L-BFGS is run
in rounds of $20$ iterations on a stratified grid of $2000$ collocation
points, which is redrawn every ten rounds so that the optimizer does not fit
a fixed set of points. Training stops when the validation loss improves by less
than $5\%$ over three rounds.
{Each $(s,\lint,m)$ configuration is trained from scratch. Future work could reduce the cost of these scans using solution bundles that learn a family of solutions over the parameters \cite{Flamant:2020bundles}, or transfer learning between nearby parameter values \cite{Desai:2022oneshot}.}

\subsection{Hard enforcement of the horizon boundary condition}
\label{sec:hardbc}

The physical boundary condition at the horizon is
Eq.~\eqref{eq:bc_horizon}. The point $z=0$ is a regular singular point of the
tilded equation, with the two Frobenius branches
\begin{equation}
  \ytilde^{\text{reg}}(z) \;=\; 1 + \frac{\atilde\,\btilde}{\ctilde}\,z
  + \mathcal{O}(z^{2}),
  \qquad
  \ytilde^{\text{sing}}(z) \;\sim\; z^{\,1-\ctilde}.
\end{equation}
The regular branch is the physical one. A na\"ive neural-network ansatz
of the form $\ytilde(z) = 1 + z\,\mathcal{N}(z)$ enforces
Eq.~\eqref{eq:bc_horizon} exactly but leaves the derivative
$\ytilde'(0) = \mathcal{N}(0)$ unconstrained. This freedom is
problematic, since it allows the network to absorb a contribution from
the singular branch $z^{1-\ctilde}$, which is precisely the unphysical
mode we seek to exclude. Since
$1-\ctilde=s-2im\gamma$ has $\mathrm{Re}(1-\ctilde)=s\ge0$, the
singular branch and its first derivative both vanish at the horizon for
$s\ge2$, so neither $\ytilde(0)$ nor $\ytilde'(0)$ detects it directly; what
distinguishes it is that $z^{1-\ctilde}$ is non-analytic at $z=0$ whenever
$m\gamma\neq0$. A parametrisation that is smooth at
the horizon by construction excludes it in principle, whereas a soft penalty
on the boundary values cannot; we return to how well this works in practice
below Eq.~\eqref{eq:ansatz}. We therefore impose {both} the boundary value
$\ytilde(0)=1$ {and}
the first Frobenius recursion~\eqref{eq:yp0} directly through the
architecture. Splitting $\ytilde = u + i v$ with $u,v:[0,1]\to\R$, we
parametrise
\begin{equation}\label{eq:ansatz}
  \ytilde(z) \;=\; 1 \;+\; \frac{\atilde\,\btilde}{\ctilde}\,z
  \;+\; z^{2}\,\bigl(\mathcal{N}_u(z;\theta) + i\,\mathcal{N}_v(z;\theta)\bigr) ,
\end{equation}
where $\mathcal{N}_u,\mathcal{N}_v$ are the two real output channels of a
single fully-connected feed-forward network whose weights and biases we denote collectively by $\theta$. The
factor $z^{2}$ guarantees that both Eq.~\eqref{eq:bc_horizon} and
Eq.~\eqref{eq:yp0} hold \emph{exactly} at every training step, for any
$\theta$. 

For $s=2$ the exclusion is not effective in practice. Near the horizon
the singular branch behaves as $z^{2-2im\gamma}$.Only the slowly varying phase $z^{-2im\gamma}$ distinguishes it from a
smooth function. We therefore solve the equation at spin weight $-|s|$ and report the result
for $|s|$, which is allowed because $k_\ell(-s)=k_\ell(s)$. At $-|s|$ the singular branch behaves as
$z^{-|s|-2im\gamma}$, diverges at the horizon, and cannot be represented by
any bounded ansatz. The change has a second effect. At integer $\ell$,
$\ytilde$ is a polynomial of degree $\ell+s$ up to $\mathcal{O}(\varepsilon)$
corrections, so lowering $s$ lowers the degree and the amplitude $|\Bgrow|$
that the network must represent.

The ansatz also has a consequence for the weighted loss introduced above.
Evaluating the residual of Eq.~\eqref{eq:system} at $z=0$ under the
constraints~\eqref{eq:bc_horizon} and \eqref{eq:yp0} gives
\begin{equation}\label{eq:resid_zero}
  R_u(0) + i R_v(0) \;=\; \ctilde\,\ytilde'(0) - \atilde\btilde\,\ytilde(0)
  \;=\; 0
\end{equation}
identically, by Eq.~\eqref{eq:indicial}. The loss weight $w(z)=1/[z(1-z)]$
diverges at $z=0$, so for a generic ansatz the weighted norm would diverge
there. The hard-enforced parametrisation is therefore required for the norm
to be finite, and not only for the reasons given above.

In the implementation, $\mathcal{N}$ is a fully-connected network {with four hidden layers} of
input dimension $1$, hidden width $128$ and output dimension $2$, with
hyperbolic-tangent activations on the hidden layers and a linear output layer. The smooth activation is essential. The loss
Eq.~\eqref{eq:loss} requires up to fourth-order derivatives of the network
output with respect to $z$, obtained by repeated reverse-mode automatic
differentiation through the $\tanh$ stack. Because $\tanh$ is $C^\infty$, the
resulting \PINN\ solution is infinitely differentiable, which is what permits
the higher-order Sobolev penalties to be defined at all. All arithmetic is
performed in double precision { (\texttt{float64}), including the network parameters, the collocation points and every automatic-differentiation pass; the exact reference solution is evaluated with \texttt{mpmath} at 50 significant digits}.

\section{From solution error to observable error}
\label{sec:amplification}
The quantity a \PINN\ minimises is the equation residual, and the quantity
reported in most \PINN\ papers is its final value. Neither is the physical
observable. The chain from one to the other has three links,
\begin{equation}\label{eq:chain}
  \mathcal{L}\;\longrightarrow\;\epsilon_y\;\longrightarrow\;
  \epsilon_A,\ \epsilon_B\;\longrightarrow\;\epsilon_k ,
\end{equation}
where $\mathcal{L}$ is the trained loss,
$\epsilon_y = \|\ytilde_{\PINN}-\ytilde_{\rm ref}\|/\|\ytilde_{\rm ref}\|$ is
the solution error, and $\epsilon_A,\epsilon_B,\epsilon_k$ are the relative
errors in the extracted coefficients and in $k_\ell = \Adec/\Bgrow$.This section quantifies the first link. 

\subsection{Training and solution profile}
\label{sec:training}

The network is trained with the weighted $H^2$ Sobolev loss, hard-enforced
boundary conditions at $z = 0$, and $400$ collocation points resampled per
epoch. Unless stated otherwise, the results of this section and of
Sections~\ref{sec:localfail}--\ref{sec:love} are for the
configuration $s=\lint=m=2$, trained with Adam alone for $2500$ epochs. The scan of
Section~\ref{sec:lmscan} uses the two-stage Adam and L-BFGS training.

\begin{figure}[H]
  \centering
  \includegraphics[width=\textwidth]{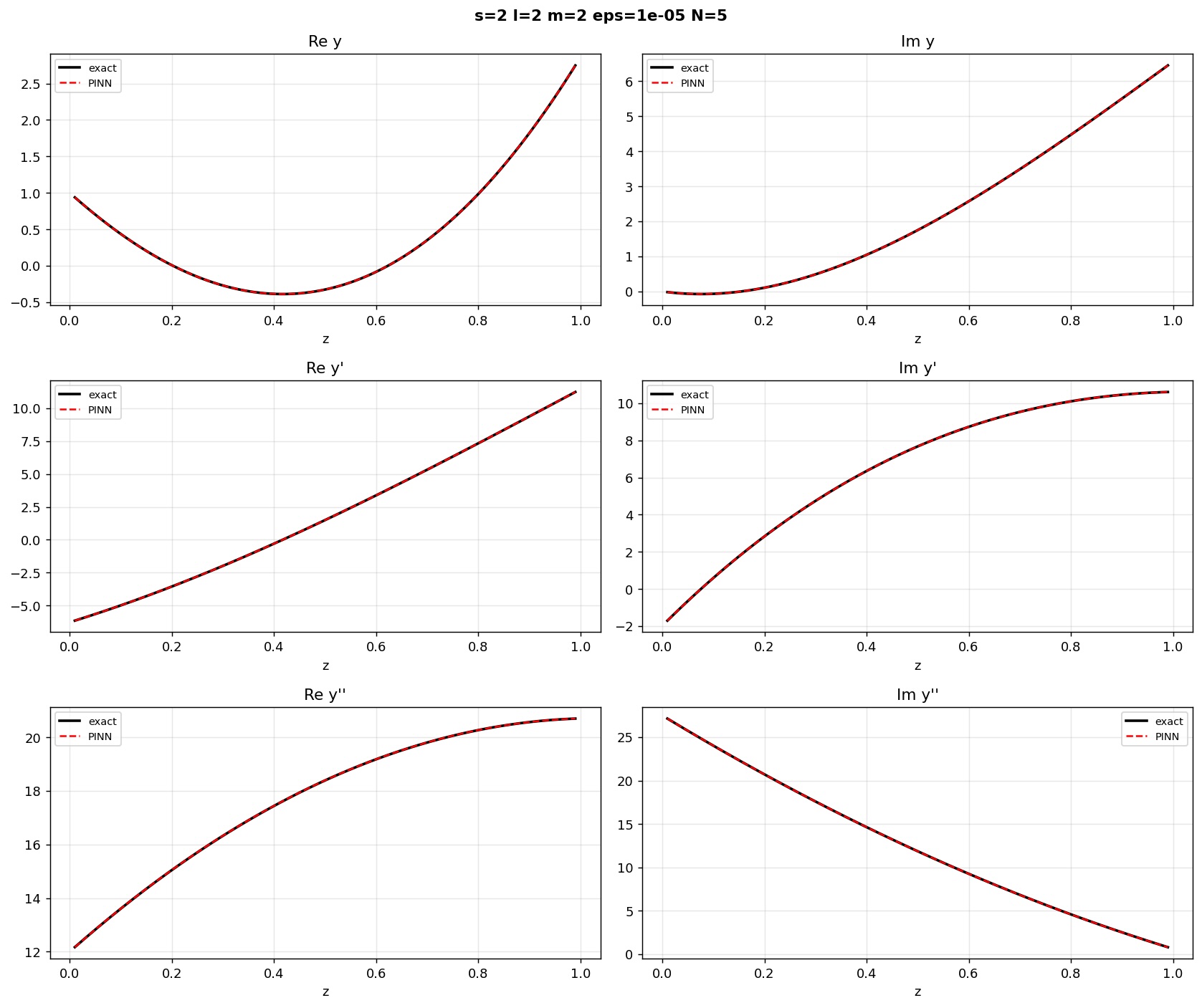}
  \caption{Trained \PINN\ output (red dashed) versus the Frobenius series
    (black solid), together with the first and second $z$-derivatives. Left
    column: $u = \operatorname{Re}\ytilde$ and right column:
    $v = \operatorname{Im}\ytilde$. Configuration $s = 2$, $\lint = 2$,
    $m = 2$, $\varepsilon = 10^{-5}$ and $2500$ epochs.}
  \label{fig:uv_profiles}
\end{figure}

\begin{figure}[H]
  \centering
  \includegraphics[width=\textwidth]{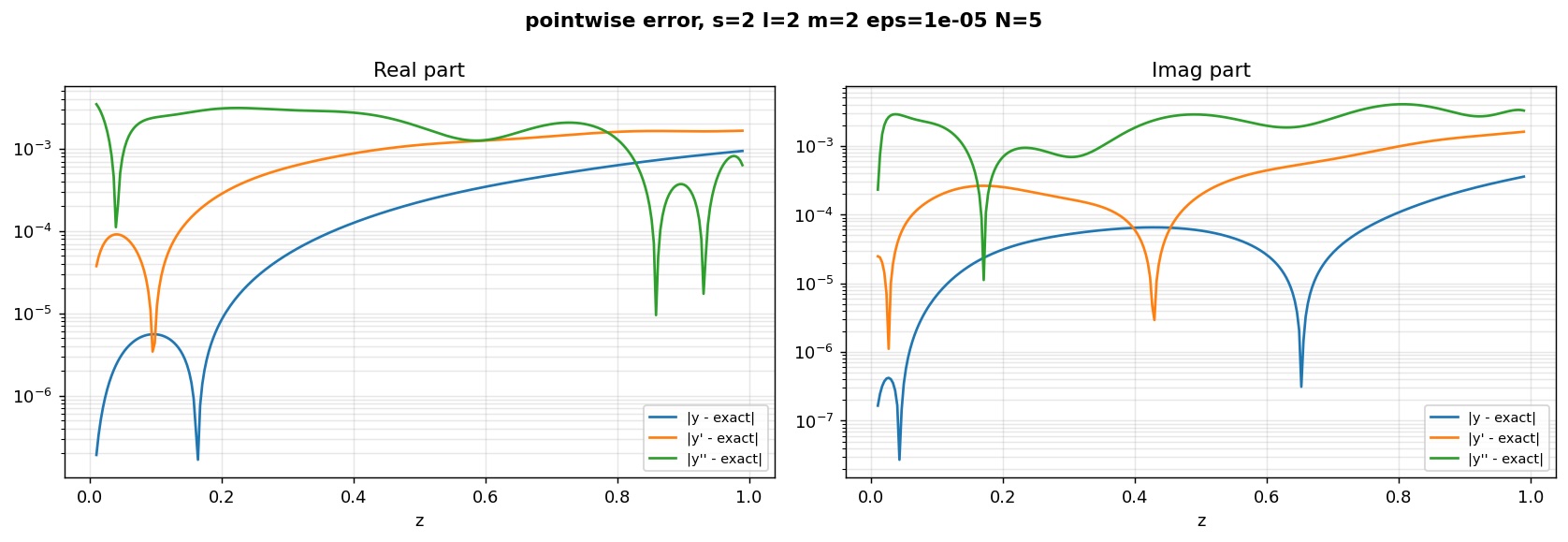}
  \caption{Pointwise error of the trained solution and of its first two
    derivatives against the exact series, on a logarithmic scale, for the
    configuration of Figure~\ref{fig:uv_profiles}. Left: real part and right:
    imaginary part.}
  \label{fig:uv_errors}
\end{figure}

Figure~\ref{fig:uv_profiles} compares the trained \PINN\ output and its first
two $z$-derivatives against the Frobenius series. The real and imaginary parts
$u$ and $v$ are reproduced closely. The first derivatives match to comparable
accuracy thanks to the $H^1$ term in the loss. The second derivatives also
track the series, with a mild endpoint deterioration as $z \to 1^-$
where the collocation density vanishes. The hierarchy
$|u - u_{\rm ex}| \lesssim |u' - u'_{\rm ex}| \lesssim |u'' - u''_{\rm ex}|$
is quantitative: at $\varepsilon=10^{-5}$ the maximum pointwise errors are
$1.0\times10^{-3}$, $2.3\times10^{-3}$ and $4.3\times10^{-3}$, each
derivative costing roughly a factor of two. The first two maxima occur at the
boundary, $z=0.99$, and the third in the interior, at $z=0.79$
(Figure~\ref{fig:uv_errors}).

\section{Extraction of physical observables}
\label{sec:extraction}

Having established a high-accuracy solver for the differential equation, we can now focus on the extraction of the physical observable and quantify the errors in the latter link of the chain~\eqref{eq:chain}.

Once the network has been trained, we read off $\Adec$ and $\Bgrow$ from the
expansion of $\ytilde$ at $z=1$, Eq.~\eqref{eq:expansion_at_1}, with the
basis $f_1,f_2$ of Eqs.~\eqref{eq:basis_frobenius}\eqref{eq:basis_frobenius2} and the \emph{real} exponent
$\Nex=\aeps+\beps-\ceps$ of {Eq.~\eqref{eq:Nex}. With this choice
Eq.~\eqref{eq:expansion_at_1} is an identity with $\Adec,\Bgrow$ exactly the
Gamma ratios~\eqref{eq:AB_gamma}, as follows from Euler's
transformation~\eqref{eq:euler} together with the Gauss connection formula.
Using the integer exponent $N$ in place of $\Nex$ breaks that identity
at $\mathcal{O}(\varepsilon)$; Section~\ref{sec:convention} quantifies the
resulting error.
However, for $\varepsilon=0$ the third parameter of $f_2$ reaches the
non-positive integer $1-N$ and the basis itself becomes ill-defined; for
$\varepsilon\neq0$ both series are well defined and we evaluate them in
arbitrary-precision arithmetic, either
from the closed forms~\eqref{eq:basis_def}\eqref{eq:basis_def2} or, equivalently, from the
recursions~\eqref{eq:basis_frobenius}\eqref{eq:basis_frobenius2}.

We implement eight extraction methods for $(\Adec, \Bgrow)$, that can be grouped in three families.

\begin{itemize}\setlength{\itemsep}{2pt}
  \item \emph{Global projection} (Methods 1, 2, 5): the solution is
    sampled on a dense subset of $(0,1)$ and projected onto $(f_1,f_2)$ by
    least squares. These average solver noise over many points and exploit the
    fact that $f_1$ and $f_2$ are well separated over the bulk of the domain
    even where their endpoint behaviours nearly coincide.
  \item \emph{Matched} (Methods 3, 4): the solution is combined with one
    or more of its derivatives at a small number of points, giving a square or
    mildly overdetermined system. These use less data and are correspondingly
    more sensitive to the conditioning of the local basis matrix.
  \item \emph{Local-derivative} (Methods 6, 7 and 8): $\Adec$ is read
    directly off the $N$-th derivative of the solution at the endpoint
    $z\to1$. This is the extraction method that Eq.~\eqref{eq:basis_def} appears to
    suggest, and Section~\ref{sec:localfail} shows that the appearance is
    misleading.
\end{itemize}

These are
\paragraph{Method 1 --- Integral residual.}
Minimize the weighted $L^2$ distance between $\ytilde_{\PINN}$ and its
decomposition,
\begin{align}
  \min_{\Adec, \Bgrow} \int_0^1 W(z)\,
  \bigl|\ytilde_{\PINN}(z) - \Adec f_1(z) - \Bgrow f_2(z)\bigr|^2 dz,
\end{align}
with $W(z) = z^{-0.3}(1-z)^{-0.4}$ emphasizing both endpoints, discretized on
$400$ points in $[0.02, 0.99]$ and solved by least squares.

\paragraph{Method 2 --- Moment matching.}
For integer moments $n = 0,\dots,3$ equate
$\int_0^1 z^n \ytilde\,dz$ against the same moments of $\Adec f_1 + \Bgrow
f_2$, giving an overdetermined system solved by least squares; the integrals
are evaluated by the trapezoidal rule on the Method-1 grid.
Explicitly,
\begin{align}
  \int_0^1 z^n\, \ytilde_{\PINN}(z)\, dz \;=\;
  \Adec \int_0^1 z^n f_1(z)\, dz + \Bgrow \int_0^1 z^n f_2(z)\, dz,
  \qquad n = 0,1,2,3.
\end{align}
The trapezoidal weights on the Method-1 grid are $t_1=(z_2-z_1)/2$,
$t_n=(z_n-z_{n-1})/2$, and $t_i=(z_{i+1}-z_{i-1})/2$ for interior indices
$i=2,\dots,n-1$. Averaging over the domain makes this method robust to
localised solver noise.

\paragraph{Method 3 --- Singular-point asymptotics.}
Since $f_1(1) = 0$ and $f_2(1) = 1$, evaluating at $z = 0.99$ and $z = 0.999$
and inverting the $2\times 2$ system gives $(\Adec, \Bgrow)$ directly. Two
function evaluations, but maximally sensitive to error in the boundary layer.

\paragraph{Method 4 --- Derivative matching.}
At $z_i \in \{0.5, 0.6, 0.7, 0.8, 0.9, 0.95, 0.99\}$, match $\ytilde$ and
$\ytilde'$ simultaneously in an overdetermined system solved by least squares.
The linear system is
\begin{align}
  \begin{pmatrix} \ytilde(z_i) \\ \ytilde'(z_i) \end{pmatrix} \;=\;
  \begin{pmatrix} f_1(z_i) & f_2(z_i) \\ f_1'(z_i) & f_2'(z_i) \end{pmatrix}
  \begin{pmatrix} \Adec \\ \Bgrow \end{pmatrix},
\end{align}
where the network derivative $\ytilde'(z_i)$ is obtained by automatic
differentiation, and $f_1', f_2'$ are computed from
Eq.~\eqref{eq:basis_def}.

\paragraph{Method 5 --- Weighted least squares near $z=1$.}
As Method 4 but value-only, at $z_i \in \{0.5,\dots,0.999\}$ with weights
$w_i = (1-z_i)^{-1/2}$ concentrating information near the endpoint.

\paragraph{Method 6 --- Direct local extraction, pointwise.}
Evaluate $\ytilde$ and $\ytilde^{(N)}$ at $z = 1-10^{-4}$ by $N$ successive
autograd passes, and read $\Adec$ and $\Bgrow$ from the local Frobenius
expansion.
Concretely,
\begin{align}\label{eq:local_naive}
  \Bgrow \;\simeq\; \ytilde_{\PINN}(1-\delta),
  \qquad
  \Adec \;\simeq\; \frac{(-1)^N}{N!}\,\ytilde_{\PINN}^{(N)}(1-\delta),
  \qquad \delta = 10^{-4}.
\end{align}
This is the most direct implementation of Eq.~\eqref{eq:basis_def} but
amplifies any high-frequency noise in the network output by the factor
$N!$, and evaluates $\ytilde^{(N)}$ at a point where the collocation
density used in training vanishes.

\paragraph{Method 7 --- Direct local extraction, Taylor fit.}
Fit a degree-$(N+3)$ polynomial in $s = z-1$ to $\ytilde$ on $15$ points
within half-width $0.02$ of $z=1$, identifying $\Bgrow$ with the constant term
and $\Adec$ with $(-1)^N c_N$.
Writing the polynomial fit as
$\ytilde_{\PINN}(z) \simeq \sum_{n=0}^{N+3} c_n\,(z-1)^n$, the identifications
are
\begin{align}
  \Bgrow \;=\; c_0,
  \qquad
  \Adec \;=\; (-1)^N c_N.
\end{align}
The $15$ points are logarithmically spaced in $|s|$ from $10^{-6}$ to
$0.02$; the polynomial degree $N+3$ gives a mildly overdetermined fit
while retaining sensitivity to the $s^N$ term.

\paragraph{Method 8 --- Endpoint-derivative extrapolation.}
Evaluate $\ytilde^{(N)}$ on a cluster
$z_k \in \{0.92,\dots,0.99\}$ inside the training domain, fit a low-degree
polynomial in $s_k = 1-z_k$ to both $\ytilde(z_k)$ and $\ytilde^{(N)}(z_k)$,
and extrapolate the fits to $s=0$; the constant term of each fit is the
extrapolated value at the boundary.
Explicitly,
\begin{align}
  \Bgrow \;\simeq\; \mathrm{poly}\bigl[\ytilde_{\PINN}(z_k)\bigr]_{s=0},
  \qquad
  \Adec \;\simeq\; \frac{(-1)^N}{N!}\,
                  \mathrm{poly}\bigl[\ytilde_{\PINN}^{(N)}(z_k)\bigr]_{s=0},
\end{align}
where each $\mathrm{poly}[\,\cdot\,]$ is a degree-$3$ polynomial in
$s_k = 1-z_k$ fit by ordinary least squares. Extrapolating from inside
the training domain avoids the point $z\to 1$ where the collocation
density vanishes, but as shown in Section~\ref{sec:localfail} the method
still suffers the exact-cancellation failure common to Methods 6, 7 and
8.

A direct comparison on the performance of these methods for the $(\ell = s = m =2)$ case is shown in Table~\ref{tab:PINNresults}.

\begin{table}[H]
    \centering
    \begin{tabular}{c|c|c|c|c}
    \toprule
    Method & \multicolumn{2}{c}{Frobenius} & 
    \multicolumn{2}{c}{PINN's} \\
     & $\operatorname{Re}k_{\ell}$ & $\operatorname{Im}k_{\ell}$ & $\operatorname{Re}k_{\ell}$ & $\operatorname{Im}k_{\ell}$ \\
    \midrule
    1 & $9.67\times 10^{-5}$ & $-3.04\times 10^{-2}$ & $6.86\times 10^{-5}$ & $-3.07\times 10^{-2}$ \\
    2 & $9.74\times 10^{-5}$ & $-3.04\times 10^{-2}$ & $1.35\times 10^{-4}$ & $-3.07\times 10^{-2}$ \\
    3 & $-1.21\times 10^{3}$ & $2.50\times 10^{3}$  & $8.75\times 10^{4}$ & $3.50\times 10^{4}$ \\
    4 & $9.68\times 10^{-5}$ & $-3.04\times 10^{-2}$ & $-1.30\times 10^{-3}$ & $-3.16\times 10^{-2}$ \\
    5 & $1.00\times 10^{-4}$ & $-3.04\times 10^{-2}$ & $2.82\times 10^{-3}$ & $-2.74\times 10^{-2}$ \\
    6 & $1.89\times 10^{-5}$ & $4.36\times 10^{-4}$ & $-1.25\times 10^{-1}$ & $-1.78\times 10^{-1}$ \\
    7 & $5.31\times 10^{8}$ & $-2.33\times 10^{8}$ & $-2.14\times 10^{6}$ & $2.15\times 10^{6}$ \\
    8 & $1.88\times 10^{-5}$ & $2.18\times 10^{-4}$ & $-1.20\times 10^{-1}$ & $-1.74\times 10^{-1}$ \\
    \bottomrule
    \end{tabular}
    \caption{Love number value $k_{\ell}$ for $\ell = m = s=2$, with $\epsilon = 1\times 10^{-3}$ and using only the ADAM optimizer. The analytical reference is $k_{\ell} = 9.63\times 10^{-5} - 3.04\times 10^{-2} i$. The \PINN\ values therefore differ from Table~\ref{tab:love_number_scan}, which reports the median over three seeds at $5000$ epochs.}
    \label{tab:PINNresults}
\end{table}

The error for both $A_{dec}$ and $B_{grow}$ is summarized in Table~\ref{tab:Errorresults}.

\begin{table}[H]
    \centering
    \begin{tabular}{c|c|c|c|c}
    \toprule
    Method & \multicolumn{2}{c}{Frobenius} & 
    \multicolumn{2}{c}{PINN's} \\
     & $|\delta A|$ & $|\delta B|$ & $|\delta A|$ & $|\delta B|$ \\
    \midrule
    1 & $1.63\times 10^{-4}$ & $5.97\times 10^{-7}$ & $5.66\times 10^{-3}$ & $5.07\times 10^{-3}$ \\
    2 & $2.90\times 10^{-4}$ & $5.31\times 10^{-7}$ & $4.57\times 10^{-3}$ & $5.04\times 10^{-3}$ \\
    3 & $9.15\times 10^{4}$ & $4.58\times 10^{-7}$  & $3.09\times 10^{6}$ & $5.19\times 10^{-3}$ \\
    4 & $1.07\times 10^{-3}$ & $2.98\times 10^{-7}$ & $5.71\times 10^{-2}$ & $5.32\times 10^{-3}$ \\
    5 & $1.32\times 10^{-3}$ & $3.67\times 10^{-7}$ & $1.37\times 10^{-1}$ & $5.18\times 10^{-3}$ \\
    6 & $1.01$ & $2.18\times 10^{-4}$ & $6.34$ & $5.40\times 10^{-3}$ \\
    7 & $1.91\times 10^{10}$ & $2.16\times 10^{-4}$ & $9.93\times 10^{7}$ & $5.20\times 10^{-3}$ \\
    8 & $1.01$ & $2.03\times 10^{-5}$ & $6.10$ & $5.18\times 10^{-3}$ \\
    \bottomrule
    \end{tabular}
    \caption{Error values for $A_{dec}$ and $B_{grow}$ for $\ell = m = s =2$, with $\epsilon = 1\times 10^{-3}$ and using only the ADAM optimizer. The analytical reference is $k_{\ell} = 9.63\times 10^{-5} - 3.04\times 10^{-2} i$.}
    \label{tab:Errorresults}
\end{table}
Every extraction method described above takes a function as input, so the same code path can be used with either the trained network or the exact series Eqs.~\eqref{eq:basis_frobenius}\eqref{eq:basis_frobenius2}. The difference between the two solutions is the
solver's contribution; anything the Frobenius series already gets wrong is a
property of the extraction method and cannot be repaired by a better solver. This
decomposition is the main diagnostic of the paper.

The series is evaluated directly at each requested point, with no
interpolation, and its truncation is constrasted against Eq.~\eqref{eq:euler}
\emph{in} \texttt{mpmath}.

\subsection{Why the local-derivative family cannot work}
\label{sec:localfail}

The local extraction methods are based on Eq.~\eqref{eq:local_naive}, which
identifies the $N$-th Taylor coefficient of $\ytilde$ about $z=1$ with
$\Adec$. The contribution from the second independent solution modifies this
coefficient. Expanding $f_2$ in powers of $(1-z)$,
\begin{equation}\label{eq:f2coeffs}
  f_2(z) \;=\; \sum_{k\ge0} c_k\,(1-z)^k,
  \qquad
  c_k \;=\; \frac{(\atilde)_k\,(\btilde)_k}{(1-\Nex)_k\;k!},
\end{equation}
the $k=N$ term occurs at the same order, $(1-z)^N$, as the leading
contribution from $f_1$. The quantity extracted by
Eq.~\eqref{eq:local_naive} is therefore
\begin{equation}\label{eq:Aeff}
  A_{\rm eff} \;=\; \Adec + \Bgrow\,c_N .
\end{equation}

The coefficient $c_N$ remains finite in the limit $\varepsilon\to0$.
As noted below Eq.~\eqref{eq:gamma}, both $(1-\Nex)_N$ and
$(\atilde)_N=(\ceps-\beps)_N$ contain a factor of order $\varepsilon$.
For the latter, this follows from
$\ceps-\beps+(\lint+s)=-\varepsilon$ and
$N=2\lint+1\ge \lint+s+1$ for $\lint\ge s$. These factors cancel, giving
$c_N=\mathcal{O}(1)$ rather than $\mathcal{O}(1/\varepsilon)$.

The limiting behavior can also be seen directly at $\varepsilon=0$. In this
limit, $\ytilde$ reduces to a polynomial of degree $\lint+s<N$, and hence
its $(1-z)^N$ Taylor coefficient vanishes. Equation~\eqref{eq:Aeff} then
implies
\begin{equation}
  A_{\rm eff}\longrightarrow 0,
  \qquad
  c_N\longrightarrow-k_\ell .
\end{equation}
Thus, as $\varepsilon\to0$, the local derivative extraction approaches zero
rather than the desired coefficient $\Adec$. The resulting relative error
therefore approaches unity. This behavior is observed for both the
neural-network and Frobenius-series solutions.

\begin{table}[H]
\centering
\begin{tabular}{lccccc}
\toprule
$\varepsilon$ & $|\Adec|$ & $|\Bgrow c_N|$ & $|c_N|$
              & $|A_{\rm eff}|/|\Adec|$ & $|c_N+k_\ell|$ \\
\midrule
$10^{-2}$ & {$0.216310$} & {$0.219416$} & {$0.029893$}
          & {$1.58\times10^{-2}$} & {$4.67\times10^{-4}$} \\
$10^{-4}$ & {$0.218198$} & {$0.218229$} & {$0.030508$}
          & {$1.60\times10^{-4}$} & {$4.88\times10^{-6}$} \\
$10^{-5}$ & {$0.218216$} & {$0.218219$} & {$0.030514$}
          & {$1.60\times10^{-5}$} & {$4.88\times10^{-7}$} \\
$10^{-8}$ & {$0.218218$} & {$0.218218$} & {$0.030514$}
          & {$1.60\times10^{-8}$} & {$4.88\times10^{-10}$} \\
\bottomrule
\end{tabular}
\caption{The quantity returned by the local-derivative extraction methods at
$s=\lint=m=2$ ($N=5$). $c_N$ is the coefficient of $(1-z)^N$ in
$f_2$, Eq.~\eqref{eq:f2coeffs}. The magnitude $|c_N|$ remains finite as
$\varepsilon\to0$, indicating that the contamination arises from
cancellation rather than divergence. The last column shows
$c_N\to-k_\ell$ linearly in $\varepsilon$, while
$|A_{\rm eff}|/|\Adec|=1.60\,\varepsilon$ {for $\varepsilon$ from $10^{-2}$ to $10^{-8}$}.}
\label{tab:contamination}
\end{table}

A separate effect enters Method 7. Since $\Nex=N+2\varepsilon$ is not an
integer for $\varepsilon\ne0$, the behavior
$f_1\sim(1-z)^{\Nex}$ introduces a branch point at $z=1$. Consequently,
$\ytilde$ is not analytic at this point, and a polynomial representation in
$(1-z)$ cannot reproduce the local behavior exactly. The amplitude of this
non-analytic contribution is $\mathcal{O}(\varepsilon)$, so this effect is
subleading compared with the coefficient contamination described by
Eq.~\eqref{eq:Aeff}.

The local extraction methods therefore provide a well-defined local
coefficient, but that coefficient contains an $\mathcal{O}(1)$ contribution
from the second independent solution. Increasing the accuracy of the
underlying numerical solution or the arithmetic precision does not remove
this systematic contribution. The global extraction methods avoid this
contamination by using information from the bulk of the domain, where the
two independent solutions remain separated.

\subsection{The exponent convention}
\label{sec:convention}

Before turning to the trained network, we quantify how the choice between the exact exponent $\Nex$ and its integer part $N$ in the field redefinition propagates into the extracted connection coefficients, and assess whether this difference produces a bias comparable in magnitude to the physical observable. The uniqueness argument below Eq.~\eqref{eq:difftilde} fixes the exponent in the field redefinition to $\Nex$, rather than to its integer part $N$. Although the two differ only at $\mathcal{O}(\varepsilon)$, this distinction can have a non-negligible effect on the extracted coefficients, as we demonstrate below.

With the integer exponent, the shifted parameters become
$\aeps - N = \atilde + 2\varepsilon$ and $\beps - N = \btilde + 2\varepsilon$,
so the solver is handed an equation displaced from
Eq.~\eqref{eq:hyper_tilde} at $\mathcal{O}(\varepsilon)$. The first basis
function then acquires a spurious factor,
\begin{equation}\label{eq:mismatch}
  (1-z)^{N} \;=\; (1-z)^{\Nex}\,(1-z)^{-2\varepsilon},
  \qquad
  (1-z)^{-2\varepsilon} \;=\; 1 - 2\varepsilon\ln(1-z)
  + \mathcal{O}(\varepsilon^2),
\end{equation}
which at $z = 0.99$ equals $1 + 9.2\,\varepsilon$. Projecting the true
solution onto this mismatched basis introduces coefficient errors at
$\mathcal{O}(\varepsilon)$.

Two consequences follow. First, with the exact exponent
Eq.~\eqref{eq:expansion_at_1} is an identity, so the extraction step
introduces no bias, and any residual error is attributable to the solver
and to the sensitivity analysed in Section~\ref{sec:amplification}.
Second, and more insidiously, an extraction error scaling as
$\mathcal{O}(\varepsilon)$ shares its functional form with the physical
vanishing of $\mathrm{Re}\,k_\ell$, so the two are easily confused.

Table~\ref{tab:convention} quantifies the difference on the Frobenius
series, with no network involved. With the integer exponent, the error
is linear in $\varepsilon$ across four orders,
$|\delta A| \simeq 0.167\,\varepsilon$; with the exact exponent, it lies
between $1.3$ and $8.4 \times 10^{-16}$ and shows no trend in
$\varepsilon$.

\begin{table}[H]
\centering
\begin{tabular}{lcccc}
\toprule
 & \multicolumn{2}{c}{integer exponent $N$}
 & \multicolumn{2}{c}{exact exponent $\Nex$} \\
\cmidrule(lr){2-3} \cmidrule(lr){4-5}
$\varepsilon$ & $|\delta A|$ & $|\delta B|$ & $|\delta A|$ & $|\delta B|$ \\
\midrule
$10^{-1}$ & $1.623\times10^{-2}$ & $4.165\times10^{-5}$
          & $2.1\times10^{-16}$  & $2.0\times10^{-16}$ \\
$10^{-2}$ & $1.667\times10^{-3}$ & $5.589\times10^{-6}$
          & $5.5\times10^{-16}$  & $2.7\times10^{-16}$ \\
$10^{-3}$ & $1.671\times10^{-4}$ & $5.785\times10^{-7}$
          & $1.3\times10^{-16}$  & $3.5\times10^{-16}$ \\
$10^{-4}$ & $1.672\times10^{-5}$ & $5.805\times10^{-8}$
          & $8.4\times10^{-16}$  & $6.3\times10^{-16}$ \\
$10^{-5}$ & $1.672\times10^{-6}$ & $5.807\times10^{-9}$
          & $3.9\times10^{-16}$  & $6.2\times10^{-17}$ \\
\bottomrule
\end{tabular}
\caption{Relative error of the integral-residual extraction applied to
the \emph{Frobenius series} at $s = \lint = m = 2$ ($N = 5$), computed in
\texttt{mpmath} at $50$ digits. Left: the superseded integer exponent
$(1-z)^{N}$. Right: the exact exponent $(1-z)^{\Nex}$. The apparent
``conditioning floor'' of the connection problem reported previously is
an artifact of the exponent mismatch~\eqref{eq:mismatch}.}
\label{tab:convention}
\end{table}

A related numerical point, easily missed, is that the hypergeometric
parameters themselves must be constructed in extended precision. The
combination $\ceps - \beps = -(\lint + s) - \varepsilon$ is a
near-integer, so in double precision its $\varepsilon$-part carries a
relative error of order $10^{-16}/\varepsilon$. The Frobenius
recursion~\eqref{eq:basis_frobenius} divides by exactly that quantity at
$n = \lint + s$, and the resulting error propagates into $\Adec$,
producing a spurious floor that grows as $1/\varepsilon$
--- $4.4 \times 10^{-11}$ at $\varepsilon = 10^{-5}$ --- independent of
the working precision of the series itself. All parameters are therefore
built in \texttt{mpmath} from exact decimal input.

\subsection{Does the Sobolev loss help?}
\label{sec:sobolev}

The $H^2$ terms in Eq.~\eqref{eq:loss} are a design choice, and we test them
directly. The same architecture is trained at the same configuration with
three settings of the Sobolev coefficients --- the default
$(\lambda_1,\lambda_2)=(0.1,0.01)$, the $H^1$ variant $(0.1,0)$, and the
plain weighted $L^2$ loss $(0,0)$ --- holding the weight $w=1/[z(1-z)]$, the
architecture, the optimiser and the collocation scheme fixed. Because the
fourth derivative is required only when $\lambda_2\neq0$ and the third only
when $\lambda_1\neq0$, the implementation takes four, three or two autograd
passes respectively, and the variants differ in cost per step as well as in
objective.

Two quantities are comparable across the three rows and one is not. The
trained loss is the value of a different functional in each case and is
listed for reference only. The solution error $\epsilon_y$ and the
coefficient error $\epsilon_A$ are the comparison.

\begin{table}[H]
\centering
\small
\begin{tabular}{lccccc}
\toprule
variant & loss$^{*}$ & $\epsilon_y$ & seed range in $\epsilon_y$
        & $\epsilon_A$ \\
\midrule
$H^2$\ \ $(0.1,\,0.01)$ & {$1.17\times10^{-4}$}
  & {$9.97\times10^{-5}$} & {$[0.71,\,1.85]\times10^{-4}$}
  & {$1.08\times10^{-4}$} \\
$H^1$\ \ $(0.1,\,0)$      & {$4.59\times10^{-3}$}
  & {$2.33\times10^{-1}$} & {$[2.22,\,3.16]\times10^{-1}$}
  & {$2.36\times10^{-1}$} \\
$L^2$\ \ $(0,\,0)$     & {$1.11\times10^{-2}$}
  & {$4.89\times10^{-1}$} & {$[4.88,\,5.81]\times10^{-1}$}
  & {$4.91\times10^{-1}$} \\
\bottomrule
\end{tabular}
\caption{Effect of the Sobolev order at $s=\lint=m=2$, $\varepsilon=10^{-3}$,
$5000$ epochs, extraction by Method 1, three seeds per row. Median over
seeds, with the seed range for $\epsilon_y$ in brackets. The loss column is
not comparable between rows.}
\label{tab:sobolev}
\end{table}

The effect is large, and it is localised in the $\lambda_2$ term. At
equal epochs the $H^2$ loss reaches $\epsilon_y\simeq1.0\times10^{-4}$, the
$H^1$ loss $2.3\times10^{-1}$ and the plain $L^2$ loss $4.9\times10^{-1}$.
Dropping the second-derivative penalty therefore costs a factor of $2300$;
dropping the first-derivative penalty as well costs only a further factor of
$2.1$. The seed spread  (the range of values for the relative error) within each variant is a factor of $2.6$, $1.4$ and
$1.2$ respectively, so both statements are far outside the noise. The
$(R'')^2$ term is not a refinement of the loss but its active ingredient,
and an $H^1$ penalty on its own is worth little more than none.

{To check that the results do not hinge on the particular choice $(\lambda_1,\lambda_2)=(0.1,0.01)$, we varied each coefficient by a factor of ten in both directions with the other held fixed, at the same configuration and with three seeds per setting (Table~\ref{tab:lambda_sens} and Fig.~\ref{fig:lambda_sens}). The solution error is insensitive to $\lambda_1$: for $\lambda_1$ from $0.01$ to $1$ it stays between $1.1$ and $1.6\times10^{-4}$, within the seed spread. It depends more on $\lambda_2$, decreasing monotonically from $3.2\times10^{-4}$ at $\lambda_2=10^{-3}$ to $3.7\times10^{-5}$ at $\lambda_2=10^{-1}$. Across all five settings $\epsilon_y$ varies by less than a factor of nine, whereas removing the $\lambda_2$ term altogether costs more than three orders of magnitude (Table~\ref{tab:sobolev}). The default weights are therefore not finely tuned; they are not even the best of the settings tried, and we keep them for consistency with the rest of the paper.}

\begin{table}[H]
\centering
\small
\begin{tabular}{ccccc}
\toprule
$\lambda_1$ & $\lambda_2$ & $\epsilon_y$ & seed range in $\epsilon_y$ & $\epsilon_A$ \\
\midrule
$0.1$  & $0.01$  & $1.22\times10^{-4}$ & $[0.76,\,2.18]\times10^{-4}$ & $1.12\times10^{-4}$ \\
$0.01$ & $0.01$  & $1.11\times10^{-4}$ & $[1.07,\,1.67]\times10^{-4}$ & $1.10\times10^{-4}$ \\
$1$    & $0.01$  & $1.57\times10^{-4}$ & $[1.25,\,2.16]\times10^{-4}$ & $1.54\times10^{-4}$ \\
$0.1$  & $0.001$ & $3.18\times10^{-4}$ & $[2.34,\,4.23]\times10^{-4}$ & $3.40\times10^{-4}$ \\
$0.1$  & $0.1$   & $3.71\times10^{-5}$ & $[3.25,\,4.67]\times10^{-5}$ & $2.73\times10^{-5}$ \\
\bottomrule
\end{tabular}
\caption{{Sensitivity to the Sobolev coefficients at $s=\lint=m=2$, $\varepsilon=10^{-3}$, $5000$ epochs, extraction by Method~1, three seeds per row; median over seeds, with the seed range for $\epsilon_y$ in brackets. The first row is the default. It was run separately from Table~\ref{tab:sobolev}, and the $\sim20\%$ difference from the corresponding row there reflects run-to-run variation.}}
\label{tab:lambda_sens}
\end{table}

\begin{figure}[H]
\centering
\includegraphics[width=\linewidth]{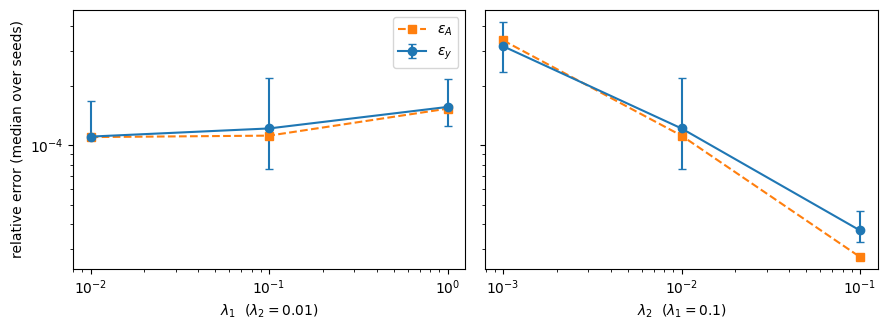}
\caption{Solution error $\epsilon_y$ (circles; bars show the range over three seeds) and coefficient error $\epsilon_A$ (squares) when varying $\lambda_1$ at $\lambda_2=0.01$ (left) and $\lambda_2$ at $\lambda_1=0.1$ (right). Same runs as Table~\ref{tab:lambda_sens}.}
\label{fig:lambda_sens}
\end{figure}

A second observation concerns the residual as a diagnostic. For the
$H^2$ variant the trained loss and the solution error are of the same order,
$1.17\times10^{-4}$ against $9.97\times10^{-5}$. For the $H^1$ and $L^2$
variants the loss is smaller than $\epsilon_y$ by factors of $51$ and $44$.
The two quantities carry different normalisations and the comparison is
therefore indicative rather than exact.

The comparison above holds the weight fixed at $w=1/[z(1-z)]$, so it
attributes to the Sobolev order whatever the two might jointly produce. To
separate them we cross the two factors, training the same architecture with
$H^2$ and with plain $L^2$, each under the divergent weight and under no
weight at all. Table~\ref{tab:crossed} gives the result.

\begin{table}[H]
\centering
\small
\begin{tabular}{llccccc}
\toprule
Sobolev & weight & loss$$ & $\epsilon_y$ & $\epsilon_A$
        & $\epsilon_y$ spread \\
\midrule
$H^2$ & $1/[z(1-z)]$ & {$5.28\times10^{-4}$}
      & {$2.15\times10^{-4}$} & {$2.06\times10^{-4}$}
      & {$[1.3,\,3.0]\times10^{-4}$} \\
$H^2$ & $1$           & {$2.88\times10^{-5}$}
      & {$4.47\times10^{-4}$} & {$4.63\times10^{-4}$}
      & {$[3.8,\,5.2]\times10^{-4}$} \\
$L^2$ & $1/[z(1-z)]$   & {$2.40\times10^{-2}$}
      & {$5.63\times10^{-1}$} & {$5.67\times10^{-1}$}
      & {$[5.1,\,6.1]\times10^{-1}$} \\
$L^2$ & $1$           & {$1.21\times10^{-3}$}
      & {$6.05\times10^{-1}$} & {$6.11\times10^{-1}$}
      & {$[5.6,\,6.5]\times10^{-1}$} \\
\bottomrule
\end{tabular}
\caption{Sobolev order crossed with endpoint weight, at
$s=\lint=m=2$, $\varepsilon=10^{-3}$, $3000$ epochs, two seeds, extraction by
Method 1. Median over seeds. The loss column is the value of a different
functional in every row and is not comparable between them.}
\label{tab:crossed}
\end{table}

The two factors are of entirely different magnitude. Dropping the
Sobolev terms costs a factor of $2600$ under the divergent weight and $1350$
under no weight. Changing the weight costs a factor of $2.1$ when the
Sobolev terms are present and $1.07$ when they are not. The endpoint weight
is therefore a second-order effect, and the small benefit it does confer
appears only in combination with the higher-derivative penalties --- which is
consistent with its role, since it emphasises precisely the region where the
higher derivatives of the solution are largest.

\subsection{Physical observable: the Love number \texorpdfstring{$k_\ell(\varepsilon)$}{kell}}
\label{sec:love}

The connection ratio $k_\ell = \Adec/\Bgrow$ is the tidal Love number. Its real and
imaginary parts encode the conservative and dissipative tidal response
respectively. Table~\ref{tab:love_number_scan} compares the trained network
against the exact Gamma ratios as the regulator is taken toward zero.

\begin{table}[H]
\centering
\small
\begin{tabular}{lccccc}
\toprule
 & \multicolumn{2}{c}{$k_\ell$ (\PINN)} & \multicolumn{2}{c}{$k_\ell$ (analytical)} & \\
\cmidrule(lr){2-3}\cmidrule(lr){4-5}
$\varepsilon$ & $\operatorname{Re}k_\ell$ & $\operatorname{Im}k_\ell$
              & $\operatorname{Re}k_\ell$ & $\operatorname{Im}k_\ell$
              & $|\Delta\operatorname{Re}k_\ell|$ \\
\midrule
$10^{-1}$ & {$+7.057\times10^{-3}$} & {$-2.1512\times10^{-2}$}
          & $+7.0495\times10^{-3}$ & $-2.15166\times10^{-2}$
          & {$7.5\times10^{-6}$} \\
$10^{-2}$ & {$+9.377\times10^{-4}$} & {$-2.9453\times10^{-2}$}
          & $+9.3337\times10^{-4}$ & $-2.94547\times10^{-2}$
          & {$4.3\times10^{-6}$} \\
$10^{-3}$ & {$+1.017\times10^{-4}$} & {$-3.0403\times10^{-2}$}
          & $+9.6321\times10^{-5}$ & $-3.04064\times10^{-2}$
          & {$5.4\times10^{-6}$} \\
$10^{-4}$ & {$+1.485\times10^{-5}$} & {$-3.0501\times10^{-2}$}
          & $+9.6628\times10^{-6}$ & $-3.05034\times10^{-2}$
          & {$5.2\times10^{-6}$}  \\
\bottomrule
\end{tabular}
\caption{Love number versus $\varepsilon$ at $s=\lint=m=2$, extraction by
Method 1, median over three seeds at $5000$ epochs. The analytical columns
are exact and convention-independent. The last column is the
\emph{absolute} error in the conservative part; it is flat while
$\mathrm{Re}\,k_\ell$ itself falls by two orders of magnitude.}
\label{tab:love_number_scan}
\end{table}

The two parts of $k_\ell$ behave in opposite ways. The dissipative part is measured. $\operatorname{Im}k_\ell$ agrees with the
exact value to a relative error of order $10^{-4}$ at every $\varepsilon$, and reproduces
the approach to its limit $-3.0514\times10^{-2}$ to four significant figures. The conservative part is only bounded. The \emph{absolute} error in
$\operatorname{Re}k_\ell$ is $5.6\times10^{-6}$ on average, between $4.3$ and $7.5\times10^{-6}$ across the four rows, and essentially independent of $\varepsilon$, while
$\operatorname{Re}k_\ell$ itself falls linearly as
$0.0967\,\varepsilon$ for $\varepsilon\lesssim10^{-2}$. The relative accuracy therefore degrades from
$10^{-3}$ at $\varepsilon=10^{-1}$ to $5\times10^{-2}$ at
$\varepsilon=10^{-3}$ and to about $0.5$ at $\varepsilon=10^{-4}$.

For each run we
measure the relative error of the extracted decaying coefficient,
\begin{equation}
  \epsilon_A \;=\; \frac{\bigl|\Adec^{\PINN}-\Adec\bigr|}{|\Adec|}\,,
  \label{eq:epsA_def}
\end{equation}
where $\Adec$ is the exact Gamma-ratio value. If this error dominates, the
error it induces in the Love number is
\begin{equation}
  |\Delta k_\ell| \;\simeq\; \epsilon_A\,\frac{|\Adec|}{|\Bgrow|}
  \;=\; \epsilon_A\,|k_\ell|\,,
  \label{eq:floor_pred}
\end{equation}
evaluated with the median $\epsilon_A$ over the three seeds and the exact
$|k_\ell|$. At $\varepsilon=10^{-4}$, for example, the three seeds give
$\epsilon_A=9.8\times10^{-5}$, $1.8\times10^{-4}$ and $8.6\times10^{-5}$, and
the median times $|k_\ell|=3.05\times10^{-2}$ gives $3.0\times10^{-6}$.
Equation~\eqref{eq:floor_pred} thus predicts $6.0$, $2.4$, $3.3$ and
$3.0\times10^{-6}$ for the four rows, against the observed $7.5$, $4.3$,
$5.4$ and $5.2\times10^{-6}$ --- agreement within a factor of $1.8$ for
$\varepsilon$ from $10^{-1}$ to $10^{-4}$. The prediction is low in every
row, as expected since it neglects the error $\epsilon_B$ in $\Bgrow$. At
$\varepsilon=10^{-4}$ the floor is already about half of
$\operatorname{Re}k_\ell$ itself, which is why the network no longer
resolves it there. The vanishing of
$\operatorname{Re}k_\ell$ is therefore not established by the network; what
the network establishes is that it is smaller than a noise floor of about $10^{-5}$, set by the accuracy $\epsilon_A$ of the extracted decaying coefficient.

It is worth emphasizing that the network does not use the Gamma-ratio
formula, and by Eq.~\eqref{eq:basis_frobenius} the basis $(f_1,f_2)$ is
generated from the coefficients of the differential equation itself. The
procedure therefore requires the local Frobenius structure at $z=1$, but not
the global connection formula --- a distinction that matters because the
former survives the confluent limit and the latter does not. We have verified
that the two constructions of $(f_1,f_2)$, the
recursions~\eqref{eq:basis_frobenius} and the closed
forms~\eqref{eq:basis_def}, agree to machine precision at every sample point
used. We stress also that the analytic values are recovered exactly by the
Gamma-ratio formula; what the \PINN\ result tests is whether a numerically
obtained solution, put through the same extraction, reproduces them.

\subsection{Scan over angular numbers \texorpdfstring{$(s, \lint, m)$}{slmscan}}
\label{sec:lmscan}

The results so far are for one configuration, $s=\lint=m=2$. Here we
check whether the method works across angular numbers and spin weights.
Table~\ref{tab:love_lms_eps3} gives the Love number for scalar ($s=0$),
vector ($s=1$) and tensor ($s=2$) perturbations with $\lint$ from $1$ to $5$
and $m=1$, and for several tensor rows with $m\neq1$. All runs use the same
network and loss as before, trained with Adam followed by L-BFGS
(Section~\ref{sec:pinn}). For $s\ge1$ we solve the equation at spin weight
$-s$, which gives the same Love number (Section~\ref{sec:hardbc}).

\begin{table}[H]
\centering
\scriptsize
\setlength{\tabcolsep}{3.5pt}
\begin{tabular}{cccccccccc}
\toprule
 & & & & & \multicolumn{2}{c}{$\operatorname{Re}k_\ell$} &
 $\operatorname{Im}k_\ell$ & \multicolumn{2}{c}{rel.\ error in $\operatorname{Im}k_\ell$}\\
\cmidrule(lr){6-7}\cmidrule(lr){9-10}
$s$ & $\lint$ & $m$ & deg & $\epsilon_y$ & \PINN & analytical & analytical &
\PINN & $\mathcal{N}\equiv0$ \\
\midrule
0 & 1 & 1 & 1 & $1.5\times10^{-6}$ & $7.71\times10^{-5}$ & $7.72\times10^{-5}$ & $-2.159\times10^{-2}$ & $4.9\times10^{-6}$ & $1.3\times10^{-4}$ \\
0 & 2 & 1 & 2 & $1.5\times10^{-7}$ & $5.39\times10^{-6}$ & $5.39\times10^{-6}$ & $-1.508\times10^{-3}$ & $5.5\times10^{-8}$ & $1.0$ \\
0 & 3 & 1 & 3 & $7.6\times10^{-7}$ & $3.55\times10^{-7}$ & $3.54\times10^{-7}$ & $-9.898\times10^{-5}$ & $6.7\times10^{-6}$ & $6.4$ \\
0 & 4 & 1 & 4 & $2.1\times10^{-7}$ & $2.27\times10^{-8}$ & $2.27\times10^{-8}$ & $-6.359\times10^{-6}$ & $5.1\times10^{-6}$ & $30$ \\
\midrule
1 & 1 & 1 & 0 & $1.6\times10^{-6}$ & $-1.54\times10^{-4}$ & $-1.54\times10^{-4}$ & $4.316\times10^{-2}$ & $6.2\times10^{-7}$ & $1.3\times10^{-6}$ \\
1 & 2 & 1 & 1 & $1.1\times10^{-6}$ & $-8.08\times10^{-6}$ & $-8.08\times10^{-6}$ & $2.261\times10^{-3}$ & $1.4\times10^{-6}$ & $8.0\times10^{-5}$ \\
1 & 3 & 1 & 2 & $2.2\times10^{-7}$ & $-4.72\times10^{-7}$ & $-4.72\times10^{-7}$ & $1.320\times10^{-4}$ & $4.9\times10^{-7}$ & $0.59$ \\
1 & 4 & 1 & 3 & $6.9\times10^{-7}$ & $-2.84\times10^{-8}$ & $-2.84\times10^{-8}$ & $7.949\times10^{-6}$ & $3.4\times10^{-7}$ & $4.8$ \\
1 & 5 & 1 & 4 & $1.2\times10^{-7}$ & $-1.74\times10^{-9}$ & $-1.74\times10^{-9}$ & $4.854\times10^{-7}$ & $7.8\times10^{-7}$ & $25$ \\
\midrule
2 & 2 & 2 & 0 & $5.4\times10^{-7}$ & $9.63\times10^{-5}$ & $9.63\times10^{-5}$ & $-3.041\times10^{-2}$ & $1.1\times10^{-9}$ & $4.8\times10^{-7}$ \\
2 & 2 & 1 & 0 & $5.4\times10^{-7}$ & $3.23\times10^{-5}$ & $3.23\times10^{-5}$ & $-9.039\times10^{-3}$ & $1.9\times10^{-8}$ & $1.3\times10^{-6}$ \\
2 & 3 & 3 & 1 & $5.1\times10^{-7}$ & $1.13\times10^{-5}$ & $1.13\times10^{-5}$ & $-3.587\times10^{-3}$ & $4.0\times10^{-9}$ & $2.0\times10^{-6}$ \\
2 & 3 & 2 & 1 & $4.3\times10^{-7}$ & $3.73\times10^{-6}$ & $3.73\times10^{-6}$ & $-1.178\times10^{-3}$ & $1.5\times10^{-8}$ & $4.1\times10^{-7}$ \\
2 & 3 & 1 & 1 & $3.7\times10^{-7}$ & $1.18\times10^{-6}$ & $1.18\times10^{-6}$ & $-3.298\times10^{-4}$ & $8.1\times10^{-8}$ & $1.6\times10^{-5}$ \\
2 & 4 & 4 & 2 & $4.8\times10^{-7}$ & $1.76\times10^{-6}$ & $1.76\times10^{-6}$ & $-5.608\times10^{-4}$ & $9.4\times10^{-9}$ & $6.7\times10^{-3}$ \\
2 & 4 & 3 & 2 & $4.1\times10^{-7}$ & $5.94\times10^{-7}$ & $5.94\times10^{-7}$ & $-1.891\times10^{-4}$ & $1.1\times10^{-8}$ & $9.9\times10^{-3}$ \\
2 & 4 & 2 & 2 & $4.8\times10^{-7}$ & $1.86\times10^{-7}$ & $1.86\times10^{-7}$ & $-5.880\times10^{-5}$ & $6.5\times10^{-9}$ & $2.8\times10^{-2}$ \\
2 & 4 & 1 & 2 & $3.1\times10^{-7}$ & $5.68\times10^{-8}$ & $5.68\times10^{-8}$ & $-1.590\times10^{-5}$ & $1.6\times10^{-7}$ & $0.15$ \\
2 & 5 & 1 & 3 & $6.2\times10^{-7}$ & $3.04\times10^{-9}$ & $3.04\times10^{-9}$ & $-8.494\times10^{-7}$ & $1.3\times10^{-6}$ & $1.5$ \\
\bottomrule
\end{tabular}
\caption{Love numbers at $\varepsilon=10^{-3}$, $a_{\rm Kerr}=0.2$, from
Method~1. Rows with $s\ge1$ are computed at spin weight $-s$, which gives the
same $k_\ell$. ``deg'' is the polynomial degree $\lint-|s|$ of $\ytilde$ at
$\varepsilon=0$ for the equation that is solved. The last column is the error
obtained with the network output set to zero, $\ytilde=1+\ytilde'(0)\,z$,
i.e.\ from the hard boundary condition alone; rows where it is already small
(deg $\le1$) are consistency checks. Single seed, $750$ Adam epochs ($1000$ for $(2,3,2)$)
followed by L-BFGS.}
\label{tab:love_lms_eps3}
\end{table}

 The last two columns of
Table~\ref{tab:love_lms_eps3} compare two errors in
$\operatorname{Im}k_\ell$: that of the trained network, and the one we would
get if the network output $\mathcal{N}=\mathcal{N}_u+i\mathcal{N}_v$ in
Eq.~\eqref{eq:ansatz} were set to zero, so that $\ytilde=1+\ytilde'(0)z$
comes only from the boundary condition at the horizon. The comparison is
needed because, at integer $\lint$, $\ytilde$ is close to a polynomial of
degree $\lint-|s|$ (column ``deg''). When this degree is $0$ or $1$, the
boundary condition already fixes almost the whole solution and the network
only adds a correction of order $\varepsilon$; these rows check that the
pipeline is consistent but do not test the solver. When the degree is $2$ or
more, the boundary condition alone gives errors between $7\times10^{-3}$ and
$30$, and the network has to learn the solution.

In the rows of degree $2$ or more the network learns
the solution accurately. The solution error is below $10^{-6}$ and the error
in $\operatorname{Im}k_\ell$ is between $10^{-8}$ and $10^{-5}$. Across the
whole table, $\operatorname{Re}k_\ell$, which at $\varepsilon=10^{-3}$ is as
small as $3\times10^{-9}$, agrees with the exact value to better than
$0.4\%$. An earlier version of this scan, trained with Adam alone at the
physical spin weight, did not converge for $\lint\ge3$ at $s=2$ and for
$\lint\ge4$ at $s=1$; with the L-BFGS stage and the change of spin weight
these rows now converge.
The equations at $s$ and $-s$ give the same
Love number, but they are not equally easy to solve. At $-s$ the solution
$\ytilde$ is a polynomial of lower degree and much smaller in size: for
$(s,\lint,m)=(2,5,1)$, for example, $|\Bgrow|$ drops from $9.3\times10^{3}$
at $s=+2$ to $12$ at $s=-2$. In addition, at $-s$ the unphysical horizon
solution diverges at $z=0$, so the network cannot absorb it
(Section~\ref{sec:hardbc}).

Figure~\ref{fig:loss_232} shows the effect on training for
$(s,\lint,m)=(2,3,2)$, solved both ways. At $s=+2$, Adam lowers the
validation loss only to about $10^{3}$ in $24$ minutes. L-BFGS then brings
it down to about $10^{-2}$, where it levels off; the solution error is
$\epsilon_y\approx0.09$ and the error in $\operatorname{Im}k_\ell$ is several
percent. At $s=-2$, Adam alone reaches $2\times10^{-8}$ in about four
minutes, and the final solution error is $4\times10^{-7}$. The plateau at
$s=+2$ is caused by the unphysical horizon solution: subtracting the multiple
of it that best fits the error of the $s=+2$ network lowers $\epsilon_y$ to
$1.6\times10^{-5}$ and the error in $\operatorname{Im}k_\ell$ to $0.26\%$.

\begin{figure}[H]
\centering
\includegraphics{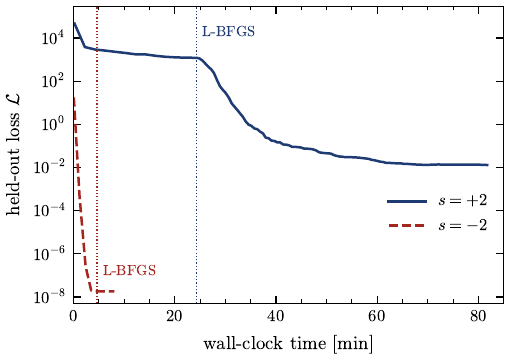}
\caption{Validation loss versus wall-clock time for $(s,\lint,m)=(2,3,2)$,
with the equation solved at $s=+2$ and at $s=-2$, which give the same
$k_\ell$. The dotted lines mark the switch from Adam to L-BFGS
(\texttt{torch.optim.LBFGS}), at about $24$\,min for $s=+2$ (blue) and about $4$\,min for $s=-2$ (red). The validation loss is the weighted $H^2$
residual on a fixed grid of $4000$ points in $z\in[0.01,0.99]$ that is not
used for training; times are for one CPU. Final solution errors:
$\epsilon_y\approx0.09$ ($s=+2$) and $4\times10^{-7}$ ($s=-2$).}
\label{fig:loss_232}
\end{figure}

\section{Discussion}
\label{sec:discussion}

We have developed and tested a \PINN s framework for computing static black-hole tidal Love numbers. The method combines a decomposition of the complex hypergeometric equation into a coupled real system, an architecture that enforces the horizon boundary condition and first Frobenius recursion exactly, and a weighted $H^2$-Sobolev loss. The response coefficients $\Adec$ and $\Bgrow$ are extracted from the trained solution using several independent methods, which are also applied to the exact Frobenius series. This provides a direct way to distinguish errors in the learned solution from limitations of the extraction procedure.

The two response coefficients present very different numerical challenges. The dominant coefficient, $\Bgrow=\ytilde(1)$, is determined directly by the boundary value and is recovered accurately for all extraction methods considered. In contrast, $\Adec$ multiplies the subdominant solution $f_1\sim(1-z)^{\Nex}$ and is therefore much more sensitive to errors in the numerical solution. The comparison of extraction methods shows that global extraction methods provide the most reliable recovery of $\Adec$, while extraction based on points very close to the singular boundary becomes increasingly sensitive as $N$ increases. The local derivative methods considered here do not recover $\Adec$ correctly. On the exact Frobenius solution they return the wrong quantity because the corresponding Taylor coefficient vanishes for the polynomial representation. This demonstrates that accurate extraction requires the extraction procedure itself to be tested independently of the neural-network solver.

The loss-function study shows that the higher-derivative terms in the Sobolev loss are essential for obtaining an accurate solution. In particular, removing the $(R'')^2$ contribution produces a dramatic increase in the solution error, while changing the endpoint weight has a comparatively modest effect (Tables~\ref{tab:sobolev} and~\ref{tab:crossed}). Thus, for the present problem, the principal improvement comes from controlling higher derivatives through the $H^2$ objective rather than from the endpoint weighting alone. This result also supersedes the corresponding conclusion of an earlier version of this work, which used a different exponent convention.

An important consequence is that the differential-equation residual should not be used by itself as a measure of solution accuracy. The loss-function comparisons show that a smaller residual does not necessarily correspond to a more accurate learned solution. The object being minimized therefore matters when interpreting \PINN \, residuals, and solution accuracy should be assessed independently whenever a reference solution is available.

The angular-number scan extends these results to $\lint\le5$ for
scalar, vector and tensor perturbations. Two changes were needed: an L-BFGS
stage after Adam, and, for $s\ge1$, solving the equation at spin weight $-s$,
which gives the same Love number and removes the singular horizon branch that
the network otherwise absorbs at $s=2$. With these, $\operatorname{Im}k_\ell$
is recovered to a relative accuracy of $10^{-5}$ or better in all rows.

The static problem provides a useful controlled setting for establishing these properties, but the framework is intended to extend to problems without closed-form solutions. Another important application is the frequency-dependent regime, in which the master equation takes the form of a confluent Heun equation. The architecture and Sobolev-loss construction can be retained, with the full $\omega$-dependent operator replacing the present residual. In this setting, the ability to impose the near-horizon behavior directly through the network architecture may be particularly useful because an exact reference solution for validating the trained network is generally unavailable.

A broader application is to modified-gravity and numerically constructed black-hole backgrounds, where the perturbation equations are not separable. In such cases, the present approach provides a possible route toward computing response functions directly from the differential equations while retaining explicit control over boundary behavior and derivative regularity.

An important special case arises for $m\gamma=0$ and $s\ge1$, where the parameter $\ceps=1-s$ takes a non-positive integer value, and the construction of
Section~\ref{sec:hyper} does not apply at the physical spin weight. At spin
weight $-s$ one has $\ceps=1+s$, so Schwarzschild and axisymmetric Kerr
perturbations may be accessible with the same method; we have not tested
this. The identity $k_\ell(-s)=k_\ell(s)$ used here holds for the static
problem, and the relation between the two spin weights has to be revisited
for $\omega\neq0$. 

Overall, our results show that \PINN s can achieve high solution accuracy for static black-hole tidal problems within a well-defined convergent regime. Achieving reliable response coefficients, however, requires the loss function, boundary conditions, neural-network solution, and extraction procedure to be considered together. The results provide a controlled demonstration of these requirements and establish a foundation for applying \PINN s to frequency-dependent and non-separable black-hole perturbation problems where conventional analytic and numerical benchmarks are no longer available.
\section*{Code availability}

The code used for all numerical results in this paper is publicly available at \url{https://github.com/RodriguezMariaJ/pinn-love-numbers}.


\section*{Acknowledgements}

We thank Luca Santoni, Adam R.\ Solomon and Malcolm Perry for illuminating discussions on
Love numbers and black hole perturbation theory. We thank Alessandro Tanzini, Giulio Bonelli, and
Cristoforo Iossa for discussions on the connection problem in
Heun-type equations. L.F.T.\ acknowledges support from the USU Howard L.\ Blood
Fellowship and the USU Physics Department Presidential Doctoral Research Fellowship. M.J.R.\ acknowledges support from the U.S.\ National Science
Foundation through grant PHY-2012036 during the initial stages of this work. M.J.R.\ also thanks the Mitchell
Family Foundation for hospitality at the Cook's Branch Workshop, USU Quantum and Gravitational Initiative
and the Instituto de F\'isica Te\'orica UAM/CSIC for hospitality
during parts of this work. We acknowledge the use of the Anthropic Claude language model as
an assistant for code and LaTeX manuscript review. All physics content,
numerical results, and scientific claims are the responsibility of
the authors.

\appendix

\section{From the Teukolsky equation to the hypergeometric form}
\label{app:teukolsky}

This appendix presents the derivation of the equations used in
Section~\ref{sec:hyper}, and fixes the relation between the hypergeometric
variable $y$ and the radial Teukolsky function.

\subsection*{Separated equations}

We start from the linear perturbation equation derived by Teukolsky
\cite{Teukolsky:1972my, Teukolsky:1973ha, Press:1973zz, Teukolsky:1974yv}.
With the ansatz
\begin{align}
    {\Phi}_{s}(t,r,\theta,\phi)=e^{-i\omega t + im\phi}R_{s}(r)S_{s}(\theta)\,,
    \qquad \omega \in \mathbb{C}, \qquad m\in \mathbb{Z}\,,
\end{align}
and $R_{s}(r) \equiv {\Delta}^{-s/2}\hat{R}_{s}(r)$ with
$\Delta = (r-r_{+})(r-r_{-})$, separability gives
\begin{align}
    &\Bigg[{\partial}_{r}\Delta{\partial}_{r}
    + \frac{\bigl(2M\omega r_{+}-\tfrac{i}{2}s(r_{+}-r_{-})-a_{\rm Kerr}m\bigr)^{2}}
           {(r-r_{+})(r_{+}-r_{-})}
    - \frac{\bigl(2M\omega r_{-} + \tfrac{i}{2}s(r_{+}-r_{-})-a_{\rm Kerr}m\bigr)^{2}}
           {(r-r_{-})(r_{+}-r_{-})} \nonumber\\
    &\qquad-{\omega}^{2}r^{2} + 2(M\omega + i s)\omega r
    + 2M\omega(2M\omega-is)-s^{2}-K_{\ell,s}\Bigg]\hat{R}_{s}=0,
    \label{eq:teuk_radial}\\[4pt]
    &\Bigg[\frac{1}{\sin{\theta}}{\partial}_{\theta}(\sin{\theta}{\partial}_{\theta})
    +(a_{\rm Kerr} \,\omega \cos{\theta}-s)^{2}
    - \frac{(m+s\cos{\theta})^{2}}{{\sin}^{2}\theta}-s^{2}+K_{\ell,s}\Bigg]
    S_{s}(\theta)=0\,.
\end{align}
In the slowly rotating limit $a_{\rm Kerr}\omega \ll 1$,
\begin{align}
    K_{\ell, s} = (\ell -s)(\ell + s +1) + s
    - a_{\rm Kerr} \,\omega \frac{2m(s^{2} + \ell(\ell + 1))}{\ell(\ell + 1)}
    + \mathcal{O}(a_{\rm Kerr}^{2} \,{\omega}^{2})\,.
\end{align}
Setting $\omega = 0$ leaves
\begin{multline}\label{eq:teuk_static}
    \Bigg[{\partial}_{r}\Delta{\partial}_{r}
    + \frac{\bigl(-\tfrac{i}{2}s(r_{+}-r_{-})-a_{\rm Kerr}  \,m\bigr)^{2}}
           {(r-r_{+})(r_{+}-r_{-})}
    - \frac{\bigl(\tfrac{i}{2}s(r_{+}-r_{-})-a_{\rm Kerr}m\bigr)^{2}}
           {(r-r_{-})(r_{+}-r_{-})}\\
           -s^{2}-K_{\ell,s}\Bigg]\hat{R}_{s}(r)=0\,,
\end{multline}
with $K_{\ell,s}=(\ell-s)(\ell+s+1)+s$.

\subsection*{Change of variable}

Equation~\eqref{eq:teuk_static} has regular singular points at $r=r_{+}$,
$r=r_{-}$ and $r=\infty$. The map
\begin{equation}\label{eq:zdef}
  z \;=\; \frac{r-r_{+}}{r-r_{-}}
\end{equation}
sends $r_{+}\!\to\!0$, $\infty\!\to\!1$ and $r_{-}\!\to\!\infty$, placing the
horizon at $z=0$ and spatial infinity at $z=1$. Writing
\begin{equation}\label{eq:Rhat_to_y}
  \hat{R}_{s}(r) \;=\; z^{\,-s/2 + i m \gamma}\,(1-z)^{\,\ell+1}\; y(z),
\end{equation}
Eq.~\eqref{eq:teuk_static} becomes the hypergeometric equation
\eqref{eq:hyper} with the parameters~\eqref{eq:abc},
where $\gamma = a_{\rm Kerr}/(r_{+}-r_{-})$.

\subsection*{Exponents}

At each singular point the two Frobenius exponents of the radial
equation~\eqref{eq:teuk_static} and of the hypergeometric equation are
related by the prefactor in Eq.~\eqref{eq:Rhat_to_y}:

\begin{table}[H]
\centering
\begin{tabular}{lccc}
\toprule
 & $r = r_{+}$ ($z=0$) & $r \to \infty$ ($z=1$) & $r = r_{-}$ ($z=\infty$)\\
\midrule
$\hat R_{s}$ & $\pm\bigl(\tfrac{s}{2} - i m\gamma\bigr)$
             & $\ell,\;-(\ell+1)$
             & --- \\
$y$          & $0,\;1-c = s-2im\gamma$
             & $0,\;c-a-b = -(2\ell+1)$
             & $a,\,b$ \\
\bottomrule
\end{tabular}
\end{table}

\noindent
The exponent differences agree in each column. At $r\to\infty$
the two radial behaviours are $r^{\ell}$, the external tidal field, and
$r^{-(\ell+1)}$, the induced response. Since
$1-z = (r_{+}-r_{-})/(r-r_{-}) \sim 1/r$, the prefactor $(1-z)^{\ell+1}$ maps
\begin{align}
  y \sim (1-z)^{0} \;&\longrightarrow\; \hat R_{s} \sim r^{-(\ell+1)}
  &&\text{(decaying: response)},\\
  y \sim (1-z)^{-(2\ell+1)} \;&\longrightarrow\; \hat R_{s} \sim r^{\ell}
  &&\text{(growing: source)} .
\end{align}

At the horizon the two exponents $\pm(s/2 - im\gamma)$ correspond to the
ingoing and outgoing solutions; the prefactor $z^{-s/2+im\gamma}$ in
Eq.~\eqref{eq:Rhat_to_y} selects the ingoing branch as the $\rho=0$ Frobenius
root of $y$ at $z=0$, which is the branch imposed by the network ansatz
\eqref{eq:ansatz}.


\section{Explicit derivatives of the residuals}
\label{app:residual_derivatives}
Applying the product rule to Eqs.~\eqref{eqn:ODEred1}--\eqref{eqn:ODEred2} gives}
\begin{align}
    R'_{u} &= (1-2z)u^{(2)} + z(1-z)u^{(3)} + P'u' + Pu^{(2)} - Q'v' - Qv^{(2)} - \alpha u' + \beta v'\,, \\
    R'_{v} &= (1-2z)v^{(2)} + z(1-z)v^{(3)} + P'v' + Pv^{(2)} + Q'u' + Qu^{(2)} - \alpha v' - \beta u'\,, \\
    R''_{u} &= -2 u^{(2)} + 2(1-2z)u^{(3)} + z(1-z)u^{(4)} + 2P'u^{(2)} + Pu^{(3)} - 2Q'v^{(2)} - Qv^{(3)} \nonumber \\
    &- \alpha u^{(2)} + \beta v^{(2)}\,,\\
    R''_{v} &= -2v^{(2)} + 2(1-2z)v^{(3)} + z(1-z)v^{(4)} + 2P'v^{(2)} + Pv^{(3)} + 2Q'u^{(2)} + Qu^{(3)} \nonumber \\
    &- \alpha v^{(2)} - \beta u^{(2)}\,,
\end{align}
Here $P'=-(a_r+b_r+1)$ and $Q'=-(a_i+b_i)$; both $P$ and $Q$ are linear in $z$, so their second derivatives vanish.

\bibliographystyle{utphys}
\bibliography{LovePinn}

\end{document}